\documentclass[aps, prl, reprint, groupedaddress, floatfix, superscriptaddress, showpacs]{revtex4-1} %showpacs preprint, reprint, twocolumn,longbibliography longbibliography,
\usepackage[colorlinks,allcolors=blue]{hyperref}
\usepackage{graphicx}
\usepackage{epstopdf}
\usepackage{dcolumn}
\usepackage{bm}
\usepackage{braket}

\newcommand{\beginsupplement}{%
    \setcounter{table}{0}
    \renewcommand{\thetable}{S\arabic{table}}%
    \setcounter{figure}{0}
    \renewcommand{\thefigure}{S\arabic{figure}}%
    \setcounter{equation}{0}
    \renewcommand{\theequation}{S\arabic{equation}}%
}

\begin{document}
\title{
	%Bond Breaking or Thermal Acceleration : \textit{ab initio} Study on Laser Induced Inertial Melting Dynamics
	%Importance of electron relaxation in lattice transformation 
    %Inertial Dynamics : Hot carrier relaxation accelerated ultrafast melting 
    %Ultrafast Carrier Relaxation Accelerates Lattice Dynamics  
    %Beyond Inertial Dynamics: Accelerated Melting by Ultrafast Carrier Relaxation   
    %Beyond Inertial Dynamics: Ultrafast Lattice Dynamics Accelerated by Carrier Relaxation
 %   Ultrafast Lattice Dynamics Accelerated by Carrier Relaxation and Phonon Generation: A Unified Picture for Laser Melting   
 %  A Unified Picture for Laser Melting: Ultrafast Lattice Dynamics Accelerated by Non-equilibrium Carrier Relaxation and Phonon Generation
 %: A Unified Picture for Laser Melting  
%Nonequilibrium Carrier Relaxation and Phonon Generation: A Unified Picture for Ultrafast Lattice Dynamics 
%Entangled Carrier-Phonon Induced Ultrafast Lattice Dynamics:  A Unified Picture  
%Entangled Carrier-Phonon Dynamics Leading to Inertial and Pauli-Drag Melting    
%Pauli-Drag Effect in Entangled Carrier-Phonon Dynamics   
%Ultrafast electron-enhanced melting and Pauli drag melting of solids
Ultrafast carrier relaxation and its Pauli drag in photo-enhanced melting of solids
% You may use “photo-induced” for “photo-enhanced” instead
}

\author{Chao Lian}
\affiliation{Beijing National Laboratory for Condensed Matter
Physics and Institute of Physics, Chinese Academy of Sciences,
Beijing, 100190, P. R. China}

\author{S. B. Zhang}\email{zhangs9@rpi.edu}
\affiliation{Department of Physics, Applied Physics, and
Astronomy, Rensselaer Polytechnic Institute, Troy, New York 12180,
USA}

\author{Sheng Meng}\email{smeng@iphy.ac.cn}
\affiliation{Beijing National Laboratory for Condensed Matter
Physics and Institute of Physics, Chinese Academy of Sciences,
Beijing, 100190, P. R. China}
\affiliation{Collaborative Innovation Center of Quantum
Matter, Beijing, 100190, P. R. China}

\date{\today}
\begin{abstract}
Ultrafast light-matter interaction is a powerful tool for the study of solids. Upon laser excitation, carrier multiplication and lattice acceleration beyond thermal velocity can occur, as a result of far-from-equilibrium carrier relaxation. The roles of electron-electron and electron-phonon scatterings are identified by first-principles dynamic simulations, from which a unified phase diagram emerges. It not only explains the experimentally-observed ``inertial" melting, but also predicts an abnormal damping by Pauli Exclusion Principle with a new perspective on ultrahigh-intensity laser applications.
\end{abstract}


\maketitle

%\section{Introduction}

%\textbf{[General description of laser induced phase transition]} It is a dream to controllably tune the interatomic potential energy surface. Compared with thermal method, laser make it fast, It is related to various fields, such as photo-chemistry.
%\begin{figure}
%    \centering
%    \includegraphics[width=1\linewidth]{inertialModelWoTitle}
%    \caption{Schematic pictures of ultrafast melting mechanism (a) `rigid' flattened PES (b) disturbed PES combined with acceleration by carrier relaxation.}
%    \label{fig:inertialModel}
%\end{figure}
%
Carrier relaxation is central to many physical, chemical, and biological processes. Not only carrier relaxation is the major source for energy loss in photovoltaic and energy conversion devices~\cite{Ulbricht2011, Bernardi2014, Jadidi2016, Brown2016a}, but it also greatly influences the lifetimes of spin and valley excitations in various emerging materials including graphene~\cite{Hwang2007, Johannsen2013, Brida2013} and transition metal dichalcogenides~\cite{Zeng2012, Bertoni2016, Mathias2016, Hao2016, Zhang2017a}. A thorough understanding is thus highly desirable, especially for gaining a control of the relaxation process at the microscopic length scale, which is an active area of nanoelectronic and optoelectronic engineering. 

In particular, laser-induced carrier dynamics has drawn great attentions because it is nonthermal, directional, and highly tunable in nature. It introduces many intriguing phenomena that have been observed experimentally, such as the ultrafast amorphization~\cite{Shank1983, Mohr-Vorobeva2011,Zalden2015, Matsubara2016, Zalden2016, Bang2016, Chen2018} and controllable order-to-order transitions~\cite{Qi2009a, Rapp2015, Iwano2017, Porer2018, Mankowsky2017, PhysRevLett.117.135501, PhysRevLett.120.185701}. It was believed that carrier relaxation takes place in a few picosecond (ps), while structural dynamics takes place in tens ps to nanoseconds. Since both processes, each assumed to be in a quasi-equilibrium, are significantly longer than the ``simultaneous" structural responses ($<$1 ps after the photoexcitation) observed in experiments~\cite{Lindenberg2005, Hillyard2007, Sciaini2009, Pardini2018}, the effects of carrier relaxation and subsequent carrier-induced dynamics for time $t<1$  ps have been purposely neglected in most theories, such as in the semiclassical models where a photoexcitation generates a sudden change in the potential energy surface (PES). The nuclei then move along this ``rigid'' adiabatic PES with a reduced transition barrier to result in the ultrafast structural changes~\cite{VanVechten1979417, VANVECHTEN1979422, Stampfli1994} (see Fig. 1). Such a picture appeared to be successful in qualitatively explaining a variety of the observations such as ultrafast amorphization~\cite{Sciaini2009,Stampfli1994,Zijlstra2008}. This is understood since first-principles excited-state dynamics simulations, which provide direct evidences for the reduced barrier and nonthermal nature of the ultrafast transitions, become available only recently~\cite{Lian2016}. 

Note that the above assumption that the structural change is triggered solely by an electronic transition, i.e., the electron and lattice degrees of freedom are completely decoupled, and the resultant PES is time-independent, i.e., ``rigid", is a serious shortcoming. As a matter of fact, it directly contradicts with recent experiments where laser induced ``inertial" lattice dynamics has been demonstrated using ultrafast X-ray and transmission electron microscopy~\cite{Lindenberg2005, Hillyard2007, Sciaini2009}. An inertial dynamics implies that the melting velocity of the ions is equal to or larger than their thermal velocity. Within the rigid PES assumption, however, this is not possible because the ions have to overcome an energy barrier before the system enters the melting phase, which effectively reduces the velocity of the ions after the phase change. To achieve an inertial dynamics in such a model thus requires the flattening of the PES, such that the ions can drift without any damping forces. As such, all the covalent bonds must be broken while all the phonon modes must be completely softened in spite that only ca. 10\% of the valence electrons are excited~\cite{Lindenberg2005, Hillyard2007, Sciaini2009}. These conclusions are in startle contrast to \textit{ab initio} simulations, showing that only acoustic phonons are significantly softened~\cite{Zijlstra2008, Lian2016}.

Without a better theory that encompasses explicitly the intertwining between electrons and phonons, further progress is essentially stalled, in spite of numerous phenomenological studies (with or without an explicit assumption on the electronic temperature)~\cite{VanVechten1979417, VANVECHTEN1979422,Stampfli1992,Stampfli1994,Zijlstra2008}. It is thus highly desirable to have a time-dependent (TD) density functional theory (DFT)-molecular dynamics (MD) study to examine the $t<1$ ps ultrafast dynamics of the excited electrons in a crystal under various excitation and temperature conditions. Such information is critically important but inaccessible by using ground-state DFT~\cite{Yabana1996, Marques2003, Castro2006, Krieger2015, Elliott2016, Brown2016a, Andrade2015}.

In this work, based on TDDFT-MD simulations, we show the fundamental importance of non-equilibrium, inherently electron-phonon-entangled carrier relaxation processes in explaining the ultrafast lattice dynamics seen by experiments. We use silicon as the prototypical system for its vastly available experimental data. However, our conclusions should be general and not limited to silicon. By analyzing the dynamic interplay between electrons (el) and phonons (ph), we show that a coupled carrier multiplication and phonon generation process during carrier relaxation is a dominant force driving ultrafast structural changes. In particular, hot electrons generated by the laser pulse are redistributed in energy through el-el and el-ph scatterings within 200 fs. The non-equilibrium and non-adiabatic process greatly accelerates the structural phase transition and is hence accountable for the inertial dynamics seen by experiments. More importantly, a new quantum phenomenon where the non-thermal melting is damped by the Pauli Exclusion Principle between the high-density carriers emerges (coined Pauli drag here), which completes the qualitative phase diagram in Fig. 1(d).  

% The flow is as follows: paragraph 1: general carrier relaxation, (2) laser-induced carrier relaxation, (3) drawback in the theory of ultrafast physics for materials, (4) the needs for our TDDFT-MD study, and (5) what we found that overturns the current understanding and new predictions (and their ramifications if possible).
\begin{figure}
	\centering
	\includegraphics[width=1.0\linewidth]{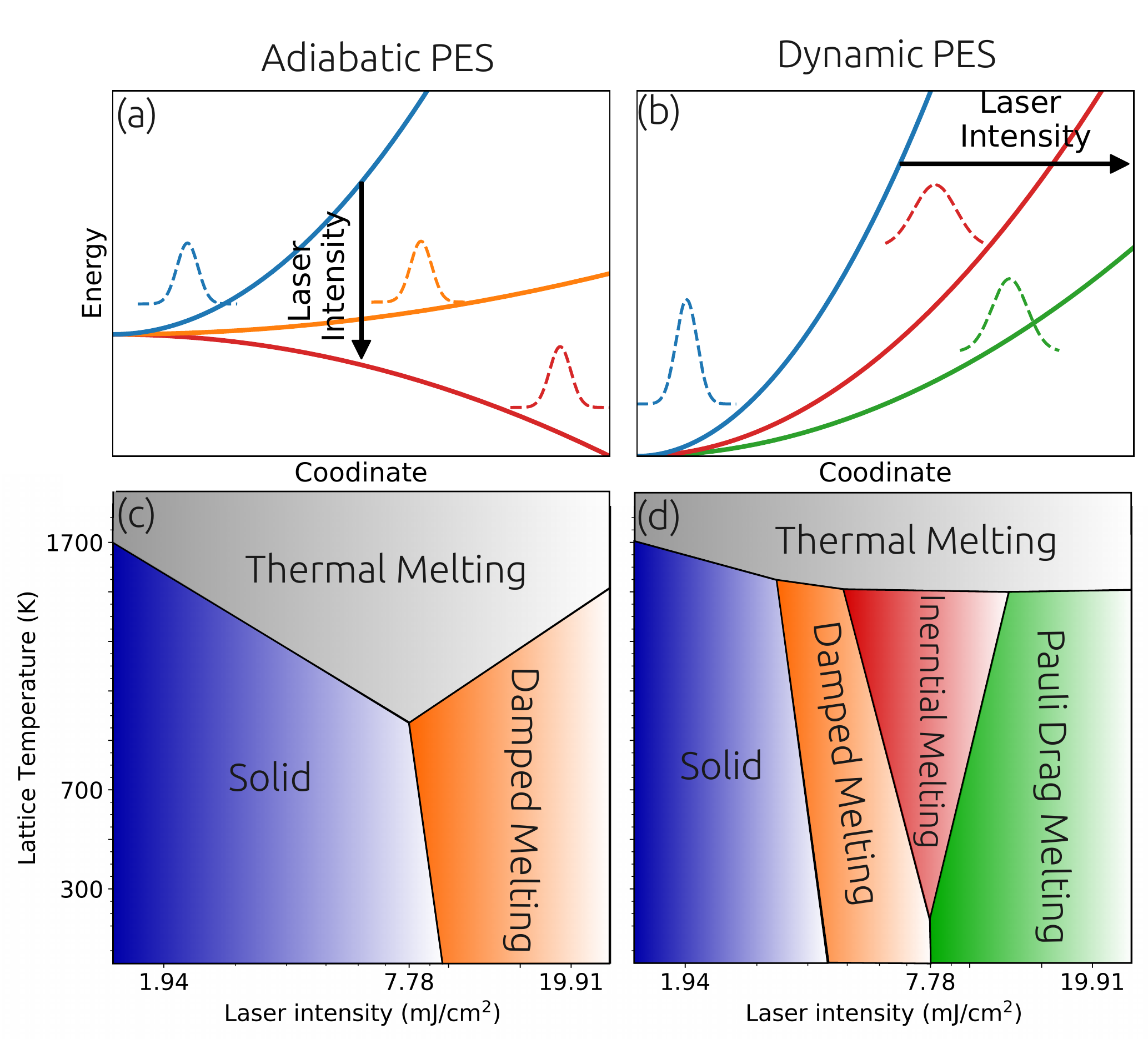}
	\caption{Schematic drawing of (a) an adiabatic and (b) a dynamic potential energy surface (PES) during ultrafast lattice dynamics. Solid lines denote the PESs, while color-coded dashed curves denote the corresponding distributions of the carriers. Schematic phase diagrams for ultrafast laser melting under (c) adiabatic PES and (d) dynamic PES, respectively. }
	\label{fig:MeltingDemo}
\end{figure}

%\section{Methods}

\begin{figure}
	\centering
\includegraphics[width=1.0\linewidth]{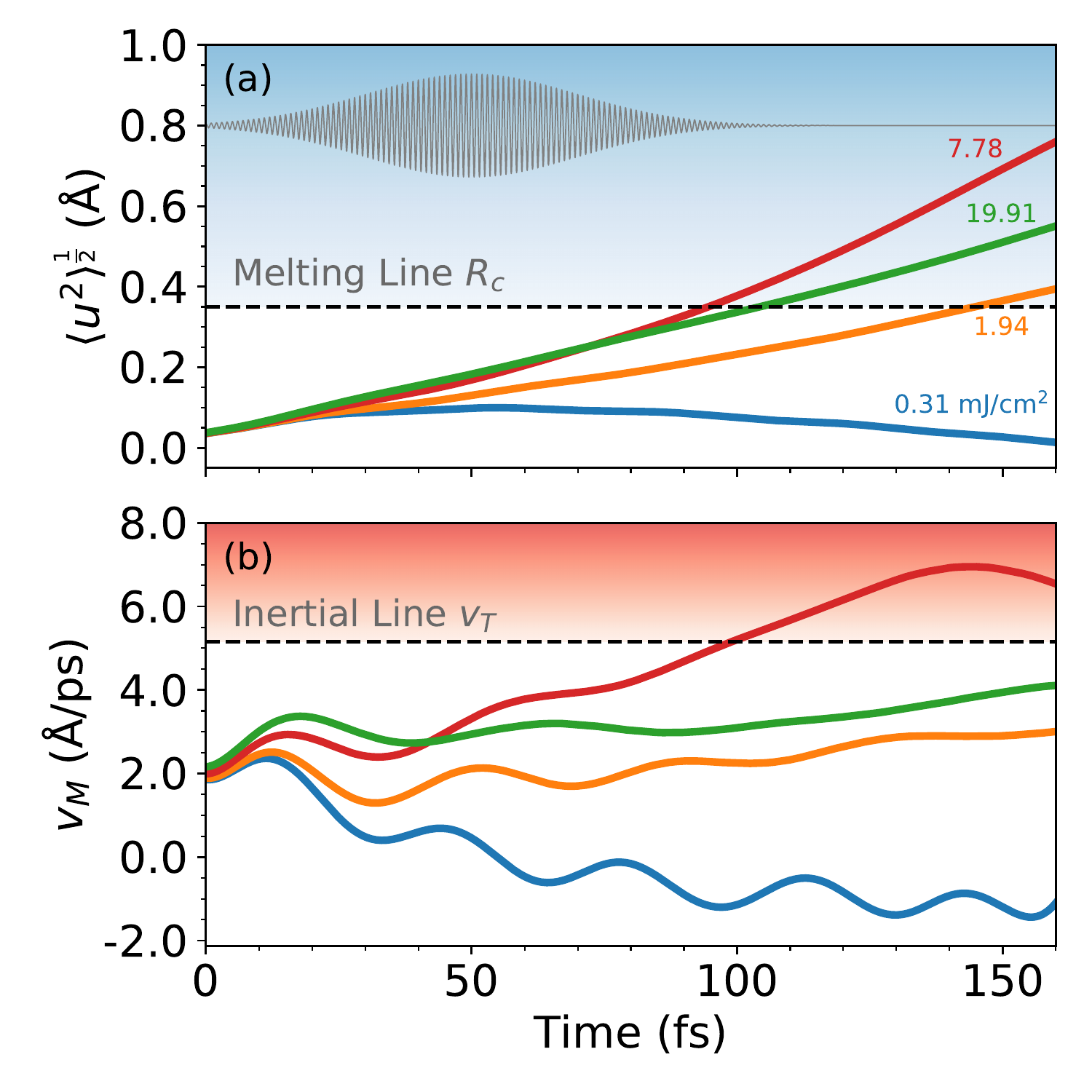}
	\caption{(a) RMSD versus time at different laser intensities with an initial temperature of 300 K. Dashed line denotes the Lindemann criterion of $R_c = 0.35$~\AA. Grey line in the inset shows the shape of the laser pulse. (b) Melting velocity $v_M$ versus time. Dashed line denotes the thermal velocity $v_T = 5.57$~\AA/ps at 300~K. 
	}
	\label{fig:inertial}
\end{figure}

The calculations were performed using a home-made real-time TDDFT code---the time dependent \textit{ab initio} package (TDAP) \cite{Meng2008, Lian2018, Lian2018b} based on the SIESTA~\cite{Ordejon1996, Soler2002, Sanchez-Portal1997}. Crystalline silicon was simulated with a periodic supercell of $64$ atoms. The Troullier-Martin pseudopotentials \cite{Troullier1991} and the adiabatic local density approximation \cite{Perdew1981, Yabana1996} for the exchange-correlation functional were used. An auxiliary real-space grid equivalent to a plane-wave cutoff of $200$~Ry was used. A Monkhosrt-Pack grid of $3\times 3 \times 3$ was used to sample the Brillouin zone. The timestep for the wavefunction evolution during the MD was $50$~attosecond for both electrons and ions. The initial atomic position and velocity were obtained from the last 1 ps of a ground-state MD simulation with a NVT ensemble. A vector gauge field \begin{equation}
\mathbf{A}(t) = -\int_0^t \mathbf{E}(t') dt'  ,
\end{equation}
was used to photoexcite the crystal. As shown in Fig.~\ref{fig:inertial}, the laser pulse took a Gaussian shape, 
\begin{equation}
\label{eq:GaussianWave}
\mathbf{E}(t)=\mathbf{E}_0\cos(\omega t) \exp\left[-\frac{(t-t_0)^2}{2\sigma^2}\right],
\end{equation}
where $E_0$ was the maximum strength of the electric field reached at $t_0 = 50$~fs, $2\sigma = 100$~fs was the pulse width, and $\omega = 4.136$~eV was the photon energy, which was the same as that used in the experiment~\cite{Harb2008}. 

Following the convention~\cite{Lindenberg2005, Lian2016}, we define the melting velocity $v_M$ as the increasing rate of the root mean square displacement (RMSD), namely, $v_M = {d \left<u^2(t)\right>^{\frac{1}{2}}}/{d t}$, and the thermal velocity as $v_T = \sqrt{3k_BT/M}$, where $M = 28.09$ is the mass of a silicon atom and $T$ is the equilibrium lattice temperature. The Lindemann criterion is used to mark the melting of the Si crystal: a melting takes place when the RMSD is larger than the critical value of $R_c = 0.35$~\AA~\cite{Lian2016}, which is roughly 15\% of the Si-Si bond length. 

%\section{Results}
%\subsection{Different Kinds of Ultrafast Melting}

\begin{table} %add [H] placement to break table across pages
\caption{\label{EC} Physical parameters of the laser pulse: $E_0$ is the maximum strength of the electric field (in V/\AA), $F$ is the fluence per pulse (in mJ/cm$^2$), and $\Delta E$ is the excitation energy (in eV/atom). Responses of silicon: $v_m^{max}$ is the maximum melting velocity (in \AA/ps), $t_m$ is the melting time (in fs) when RMSD reaches the Linderman criterion, $\tilde{T}_{max}$ is the maximum nominal lattice temperature (in K), $n^{\mathrm{laser}}_e$ and $n_e^{\mathrm{multi}}$ are the numbers of carriers, which are either laser-induced or originated from carrier multiplication at 150 fs, respectively (in $0.01$~e/atom), and $T_e$ is the fitted electronic temperature (in eV).}
    \begin{ruledtabular}
        \begin{tabular}{ccccclcccccc}
        	    $E_0$  & $F$    &  $\Delta E$   &  $v_M^{max}$  &      $t_m$      & $\tilde{T}_{max}$      &     $n_e$     & $n_e^{\mathrm{multi}}$ &    $T_e$     &  \\ \hline
        	    0.026   &0.31   &     0.73      &     2.01      &    $\infty$     & 370.4          &     16.9      &         -0.23          &     1.6      &  \\
        	    0.064  &1.94    &     1.25      &     2.73      &     157.05      & 378.8          &     29.0      &          0.42          &     2.0      &  \\
        	    0.116 &7.78     &     2.39      & \textbf{6.31} & \textbf{101.75} & \textbf{745.5} &     50.7      &     \textbf{3.50}      &     3.2      &  \\
        	\textbf{0.206} & \textbf{19.91}& \textbf{4.26} &     3.40      &     115.80      & 421.2          & \textbf{69.0} &          1.48          & \textbf{4.4} &
        \end{tabular}
    \end{ruledtabular}
\end{table}

Figure~\ref{fig:inertial} and Table~\ref{EC} show the results under various excitation conditions: laser field strength is $E_0 = 0.026, 0.064, 0.116, 0.206$~V/\AA\ and laser fluence is $F=$ 0.31, 1.94, 7.78, and 19.91 mJ/cm$^2$, respectively, from which four types of lattice dynamics can be identified. (i) No melting: at a lowest fluence $F=0.31$~mJ/cm$^2$, the RMSD, as expected, only oscillates around zero. (ii) Damped melting: with $F$ increases to 1.94~mJ/cm$^2$, the system starts to melt with its RMSD surpassing the Lindemann criterion at $t_m=157$~fs, while the maximum melting velocity $v_M^{max} = 2.73$~\AA/ps is still far below the thermal velocity, $v_T = 5.57$~\AA/ps. (iii) Inertial melting: with $F$ further increases to $7.78$~mJ/cm$^2$, the melting time $t_m$ is noticeably shortened. Here, the melting is accelerated with $v_M$ exceeding $v_T$ at $99$~fs. (iv) Recurrence of damped melting: with $F$ further increases to $19.91$~mJ/cm$^2$, the system goes back surprisingly to the damped melting with a maximum velocity $v_M = 3.40$~\AA/ps. In our simulation, a deceleration of $v_M$ is generally observed for $F>9.77$~mJ/cm$^2$ (Fig.~S1). 

\begin{figure}
\centering
\includegraphics[width=1.0\linewidth]{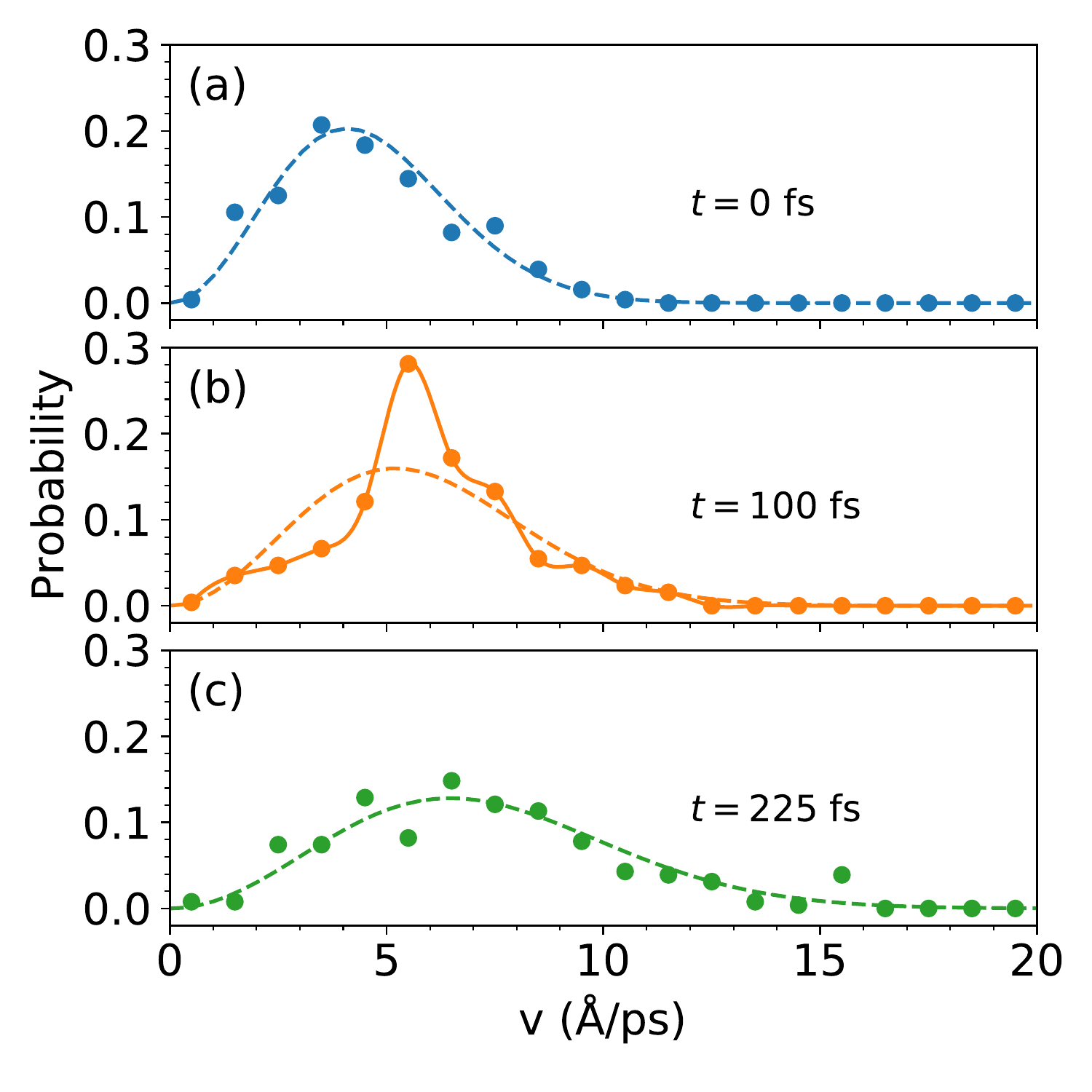}
\caption{Ionic velocity distributions for laser fluence $F=7.78$~mJ/cm$^2$ at (a) $t=0$~fs, (b) $100$~fs, and (c) $225$~fs. Dashed lines denote the Maxwell distributions with nominal lattice temperature $\tilde{T}=283, 456, 707$~K, respectively.}
\label{fig:ionVelDist}
\end{figure}

Based on the results above, we can draw a schematic $T-I$ phase diagram [Fig.~\ref{fig:MeltingDemo}(d)] for Si crystal, where $I$ is the laser intensity. At a low initial $T$ and a low $I$, no melting takes place. With an increase of $I$, nonthermal melting occurs. Within this regime, the inertial melting takes place at a medium strength $I$. With $I$ further increases, however, the system goes to a new regime, termed Pauli drag melting, to be extensively discussed below. With $T$ increases, on the other hand, thermal effect becomes dominant, leading to conventional thermal melting. 

%\subsection{The Effect of Carrier Relaxation}

\begin{figure*}
    \centering
        \includegraphics[width=1.0\linewidth]{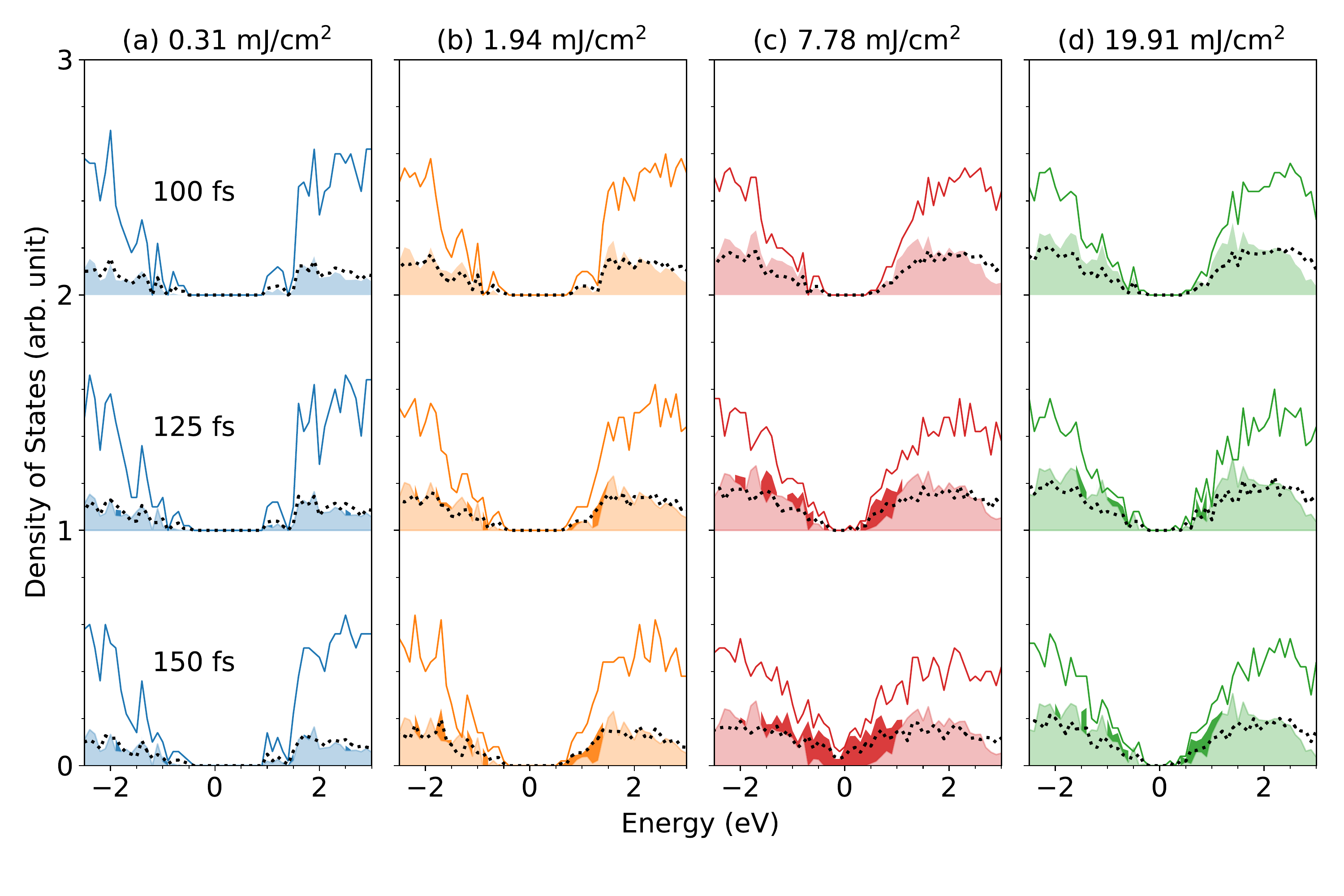}
    \caption{\label{fig:relaxDemo} Time snapshots of carrier distributions over energy (in eV) at different laser intensities. Zero in horizontal axis is the Fermi energy. Solid lines denote electronic densities of states, filled regions denote populations of carriers, with darker regions highlight the difference between the population at the moment and that at $t=$100 fs, and dotted lines denote populations of carriers with electronic temperature $T_e$ = (a) 1.6, (b) 2.0, (c) 3.2, and (d) 4.4~eV.}
   
\end{figure*}

\begin{figure}
	\centering
	\includegraphics[width=1.0\linewidth]{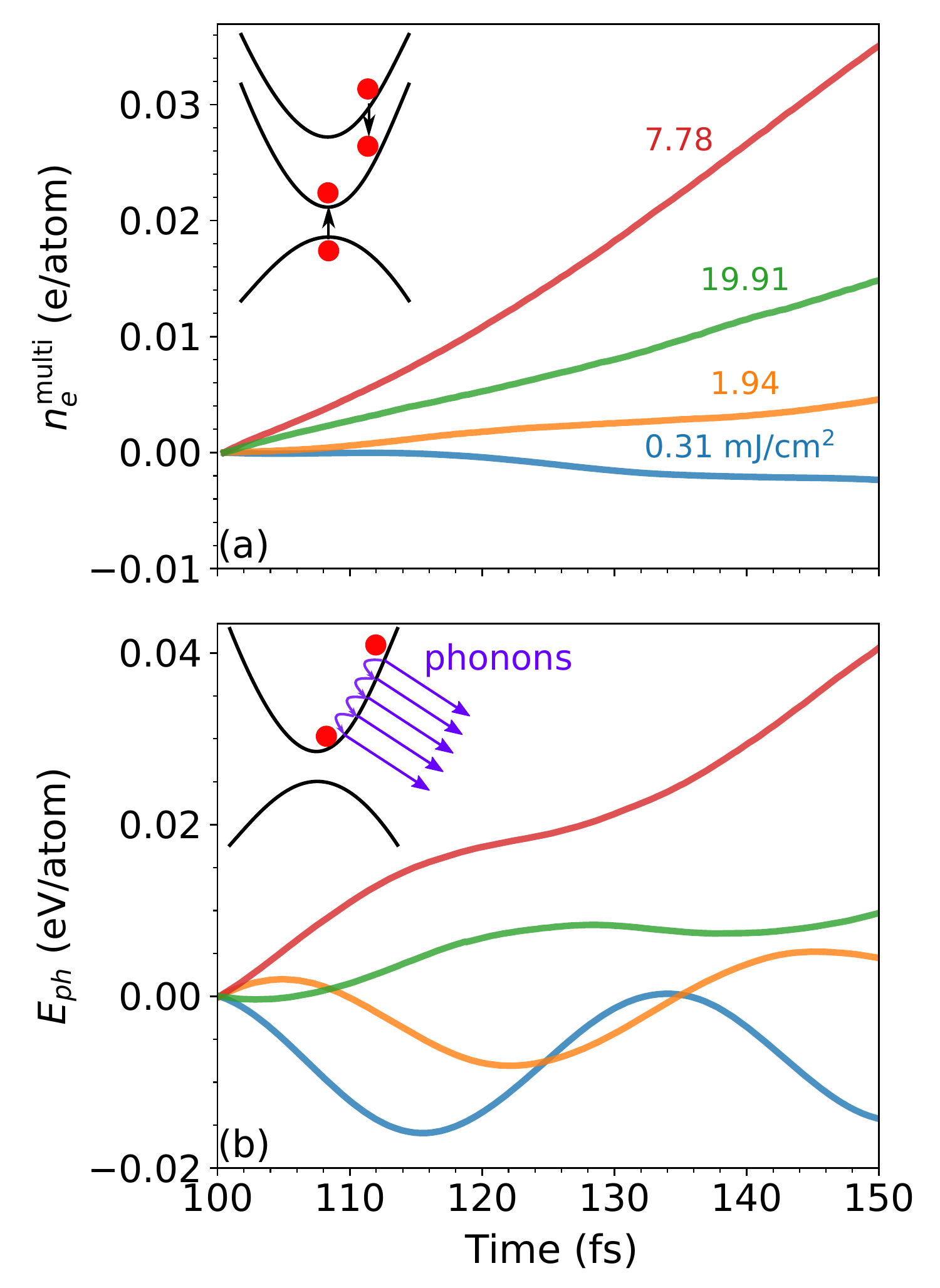}
	\caption{\label{fig5} Carrier-carrier and carrier-phonon scatterings (as indicated in the insets). (a) Carrier density due to carrier multiplication $n_e^{\mathrm{multi}}(t)$ and (b) phonon energy versus time. In (b), energy at 100 fs is set to zero.}
 
\end{figure}

A rigid adiabatic PES model has difficulty to explain the rich physics in Fig.~\ref{fig:MeltingDemo}(d), in particular, the unexpected deceleration at a high laser intensity, which would require the melting barrier first decreases but then increases with laser intensity. In principle, a deceleration of $v_M$ is possible, provided that the high-lying conduction band states occupied under $F>9.77$~mJ/cm$^2$ are predominantly the bonding states. From a crystal orbital Hamilton population (COHP) analysis~\cite{Dronskowski1993} of TDDFT-MD trajectory (Fig.~S2), we see that all the conduction bands are in fact anti-bonding states, which suggests that an increase in the laser intensity should only weaken the Si-Si bonds and subsequently a lowering of the melting barrier. Thus, the observed deceleration of $v_M$ cannot be a consequence of an abnormal barrier change.

Our TDDFT-MD simulations offer hints to the problem. Consider the nominal lattice temperature after photoexcitation, $\tilde{T}(t) = \sum_i v^2_i(t)/2M$ where $v_i$ is the ionic velocity of the $i$th atom. At $F=7.78$~mJ/cm$^2$ where the inertial dynamics was observed, the lattice is heated considerably from an initial $T=300$~K to a $\tilde{T}_{max}=745.5$~K (see Table~\ref{EC}). At other laser intensities (either lower or higher), in contrast, no such inertial dynamics was observed. Instead, $\tilde{T}(t)$ oscillates around 300~K with the maximum $\tilde{T}_{max}\le 421$~K (Table~\ref{EC}). Note that $\tilde{T}$ here is different from the equilibrium temperature $T$.

For its uniqueness, let us now examine $F=7.78$~mJ/cm$^2$. Figure~\ref{fig:ionVelDist} shows that at $t=0$~fs the ionic velocities of individual atoms adopt an equilibrium Maxwell distribution. After laser irradiation, however, a clear derivation from the equilibrium distribution is observed: at $t=100$~fs, the peak is shifted towards a higher velocity and its width becomes much narrower [see Fig.~\ref{fig:ionVelDist}(b)]. At $t=225$~fs, the Maxwell distribution approximately recovers but with a larger standard deviation. In other words, the system evolves from an equilibrium state to a nonequilibrium one, but then rapidly recovers to a quasiequilibrium state. During the process, no equilibrium $T$ can be defined. In order to evaluate the average kinetic energy of the ions, we use the nominal temperature $\tilde{T}(t)$. 

We find that nonequilibrium, ultrafast carrier relaxation is the reason for the increase in $\tilde{T}(t)$. Below, we focus on two major carrier relaxation mechanisms. The first one is carrier multiplication. As an excited carrier is relaxed to a lower energy state, another electron in its ground state can be excited across the band gap. Carrier multiplication represents the net effect of an Auger recombination and impact ionization. In this process, the energy of the electronic subsystem is preserved, while the number of the carriers is increased. The second mechanism is carrier-phonon scattering. When a carrier is scattered between states at different momenta, phonons are emitted or absorbed to conserve the total momentum.  In this process, energy is transferred from the electronic subsystem to the lattice, which increases ionic kinetic energy. 

Figure~\ref{fig:relaxDemo} shows the energy distribution of the excited carriers. At $F=0.31$~mJ/cm$^2$, the distribution barely changes with time. As such, it is well described by a Fermi-Dirac (FD) distribution (i.e., the dotted line in Fig.~\ref{fig:relaxDemo}, albeit at a high electronic temperature $T_e = 1.6$~eV $= 1.9\times10^4$~K). At $F=1.94$ and 7.78~mJ/cm$^2$, a significant deviation from the FD distribution is observed. It signals the intrinsic difficulty in assigning an explicit $T_e$ to such a nonadiabatic dynamic system~\cite{Zijlstra2008}. 
%Noticeable carrier relaxations are also observed with the relaxation of the excited electrons faster than that of the excited holes. 
%As a result of the relaxation, carrier distribution approaches the FD distribution when $t$ is increased from 100 to 150 fs. 
A higher laser intensity corresponds to a higher (fitted) $T_e$ (cf. Table~\ref{EC}). Despite that, the carriers are not in equilibrium as evidenced by significant derivation from the FD distribution for $|E|>2.5$~eV.  Meanwhile, the band gap is closing as a result of the carrier relaxation. A decrease in the gap in turn lowers the threshold for the carrier multiplication. In other words,  one has a self-amplified process due to the interplay between carrier multiplication and gap closure. 
%More detailed analysis on this can be found in Ref.~\cite{SM}.

To quantify the effect of carrier multiplication, we define the net increase in carriers during the relaxation process as $n_e^{\mathrm{multi}}(t)=n_e(t)-n_e^{\mathrm{laser}}$, where $n_e(t)$ is the total number of carriers at time $t$ and $n_e^{\mathrm{laser}}$ is the number of carriers generated by the laser pulse right after the laser field has diminished. In our discussion, $n_e^{\mathrm{laser}} = n_e(t=100~\mathrm{fs})$. Figure~\ref{fig5}(a) and Tab.~\ref{EC} show that $n_e^{\mathrm{laser}}$ increases monotonically with $F$, while $n_e^{\mathrm{multi}}$ shows an abnormal decrease when $F$ increases from  7.78 to 19.91~mJ/cm$^2$. Hence, it is $n_e^{\mathrm{multi}}$, not $n_e^{\mathrm{laser}}$, that shares the same trend with $v_M$.

To examine the dynamic effect of phonons, we define the phonon energy as $E_{ph}(t) = E_{tot}(t) - E_{KS}(t)$, where $E_{tot}(t)$ is the total energy of the system, and $E_{KS}(t)$ is the Kohn-Sham energy of the electronic subsystem. Figure~\ref{fig5} shows that, at a low fluence, $E_{ph}(t)$ and $n_e^{\mathrm{multi}}(t)$ are seemingly unrelated. However, at a high fluence, the two become correlated, evidenced by the fact that $E_{ph}(t)$ increases with $n_e^{\mathrm{multi}}(t)$, and at $F=$7.78~mJ/cm$^2$, both $E_{ph}$ and $n_e^{\mathrm{multi}}$ reach their maximum. This observation is an indication that carrier multiplication and phonon generation are an entangled physical process. As a matter of fact, the carriers and phonons may even form dynamic polarons, but a further analysis would be beyond the scope of the current work.

The above analysis also offers a physical explanation to the inability to further increase $v_M$ at $F=$19.91~mJ/cm$^2$, namely, a suppression of the lattice dynamics. It comes about because of the blocking of the effective carrier relaxation pathways: when a large number of electrons populate the conduction band under a strong laser illumination, fewer empty states are available for the relaxation of the higher-energy hot electrons. In essence, this happens because electrons are fermions; Pauli Exclusion Principle prevents them from taking the already-occupied electronic states, whereby leading to a damped carrier relaxation and lattice dynamics. Hence, we will term such an effect a Pauli drag effect. 

Strictly speaking, carrier relaxation affects both PES and $\tilde{T}$. However, because the excitation energy $\Delta E \sim 3$~eV/atom is much larger than $E_{ph}(t) \sim 0.04$~eV/atom (Table~\ref{EC}), the decrease in electronic energy, caused by a direct carrier-phonon scattering, will be small. Instead, the effect of carrier multiplication will be more pronounced. With more carriers occupying the low-energy bands, the PES is in turn significantly altered to facilitate phonon scattering with low-energy carriers. This coupled process explains the correlated carrier-phonon dynamics and enhanced phonon generation.

Note that this work focuses on the silent physics in initial stage ($t<200$~fs) of laser excitation when the decoherence of lattice vibration is still insignificant. Phonon-phonon scattering become important after picoseconds, leading to an equilibration between electrons and phonons. The energy due to laser irradiation will dissipate into the environment, or cause an irreversible damage to the material such as destruction or ablation. We expect that the abnormal deceleration of melting at high laser intensity, i.e., the Pauli drag effect, will be convoluted with these processes at longer timescales. Regardless, the predicted Pauli drag effect should be readily measured in an ultrafast X-ray or electron diffraction experiment: the indication of the effect would be a deterred melting at an increased excitation fluence.  

%\section{Conclusion}
In conclusion, we have studied, using a TDDFT-MD approach and Si as a porotype, the ultrafast lattice dynamics under laser excitation. Our results reveal the physics at high excitation intensities that consists of both an enhanced and a decelerated melting regime, driven by an entangled nonequilibrium carrier multiplication and phonon generation process. The accelerated process resolves the longtime mystery surrounding the inertial dynamics observed by experiment, while at an even higher laser intensity, the decelerated melting phenomenon takes over, as a result of the Pauli Exclusion between high density carriers. Since inertial melting has been a roadblock to the development of ultrahigh-power laser materials and devices, the identification of the Pauli drag regime offers potentially a completely different perspective in fabricating materials and engineering devices that survive intense lasers for unprecedented applications.
   
%\section{Acknowledgment}
CL and SM acknowledge partial financial supports from MOST (grants 2016YFA0300902 and 2015CB921001), NSFC (grants 11774396 and 11474328), and CAS (XDB07030100). SBZ acknowledges the support by US Department of Energy under Grant No. DE-SC0002623.

\beginsupplement
\section{Supplementary Materials}

\subsection{Carrier Scattering in TDDFT}
    	
We evaluate the state-to-state transition probabilities between TDKS orbitals during time evolution:
\begin{equation}
\label{projection}
P_{nn'\mathbf{k}} = \left|C_{nn'\mathbf{k}} \right|^2 = \left|\braket{v_{n\mathbf{k}}|S_k|u_{n'\mathbf{k}}}\right|^2,
\end{equation}
where $S_\mathbf{k}$ is the overlap matrix,  $u_{n'\mathbf{k}}$ is the time dependent Kohn-Sham (TDKS) orbitals and $\ket{v_{n\mathbf{k}}}$ is the adiabatic basis satisfying
\begin{equation}
\label{adiabaticBasis}
H_\mathbf{k}\ket{v_{n\mathbf{k}}(\mathbf{r})} = E_{n\mathbf{k}} S_\mathbf{k}\ket{v_{n\mathbf{k}}(\mathbf{r})}.
\end{equation}
where $H_\mathbf{k}$ is the Hamiltonian.
The population $\mathcal{q}_{n\mathbf{k}}$ of the adiabatic state $n\mathbf{k}$ is thus projected from the TDKS orbitals at a given time as:
\begin{equation}
\label{eq:excitation}
\mathcal{q}_{n\mathbf{k}} =  \sum_{n'\in n_{\mathbf{k},occ}} q_{n'\mathbf{k}} P_{nn'\mathbf{k}},
\end{equation}
where $n_{\mathbf{k},occ}$ is the occupied state at $\mathbf{k}$ point.
The number of excited electrons $n(t)$ is calculated as,
\begin{equation}
\label{eq:ExcitedElectrons}
n(t) = \sum_{unocc} \mathcal{q}_{n\mathbf{k}}(t),
\end{equation}

%\subsection{Describe the Scattering in TDDFT}
The scattering of carriers are described by the changes of $\mathcal{q}_{n\mathbf{k}}$ as a function of time. The increase of $\mathcal{q}_{n\mathbf{k}}$ occurs together with the decrease of $\mathcal{q}_{m\mathbf{k}}$, which represents the transition from adiabatic state $m$ to $n$. The transition is confined within the same $\mathbf{k}$, since $H_\mathbf{k}$ is independent subspace for each $\mathbf{k}$. Thus, only intra-$\mathbf{k}$ scattering is allowed in the simulation. This limitation requires a supercell to accurately describe the scattering process. With a infinite large supercell, all inter-$\mathbf{k}$ scatterings in unit cell are converted to intra-$\mathbf{k}$ scatterings, due to the band folding. 

%The density of states is a qualitative evaluation of the accuracy. The 

%\newpage
\begin{figure}
    \centering
    \includegraphics[width=1.0\linewidth]{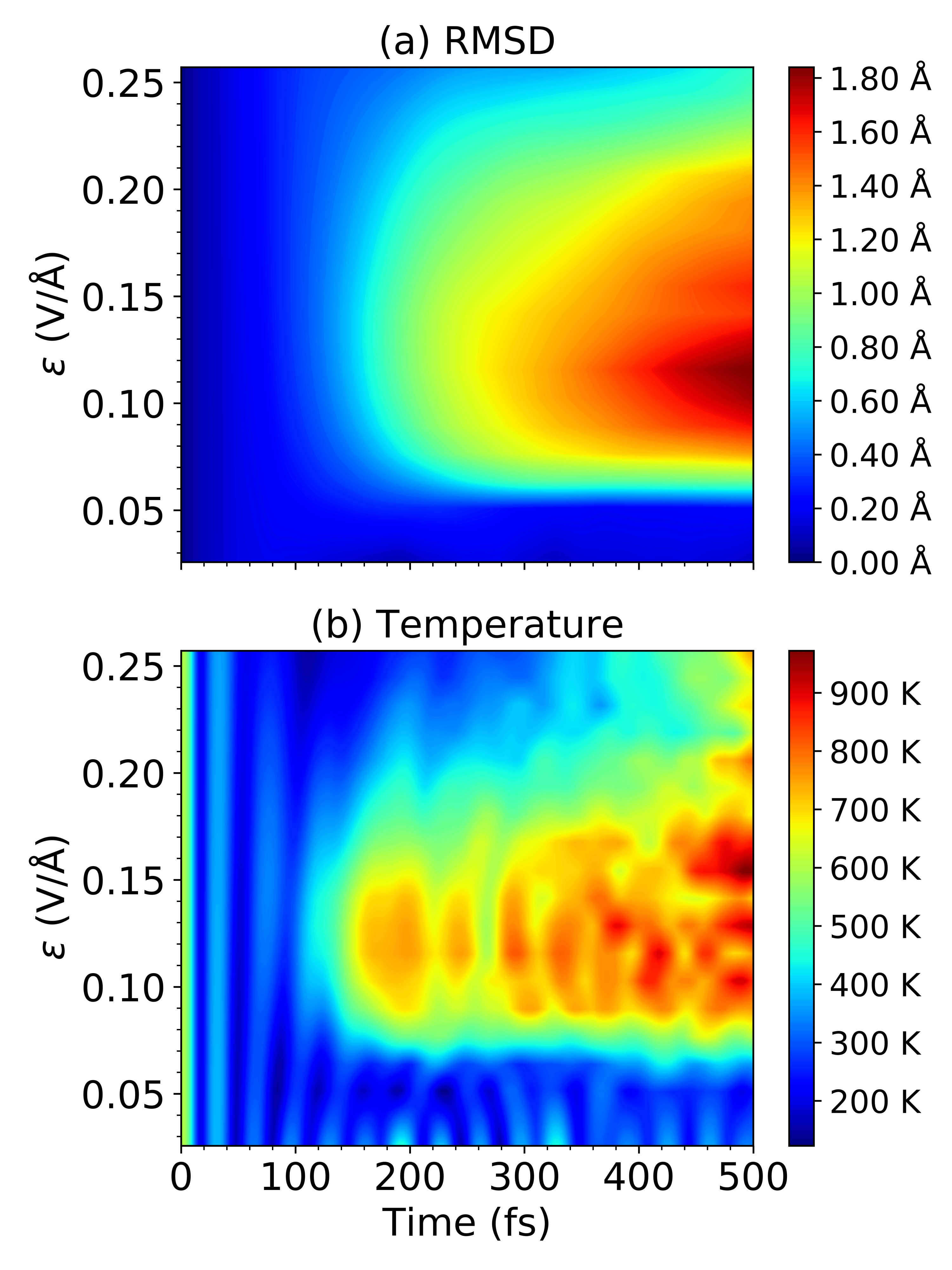}
    \caption{(a) RMSD and (b) temperature contour maps as a function of time. }
    \label{fig:RMSDContour}
\end{figure}

%\newpage
\begin{figure}
    \centering
    \includegraphics[width=1.0\linewidth]{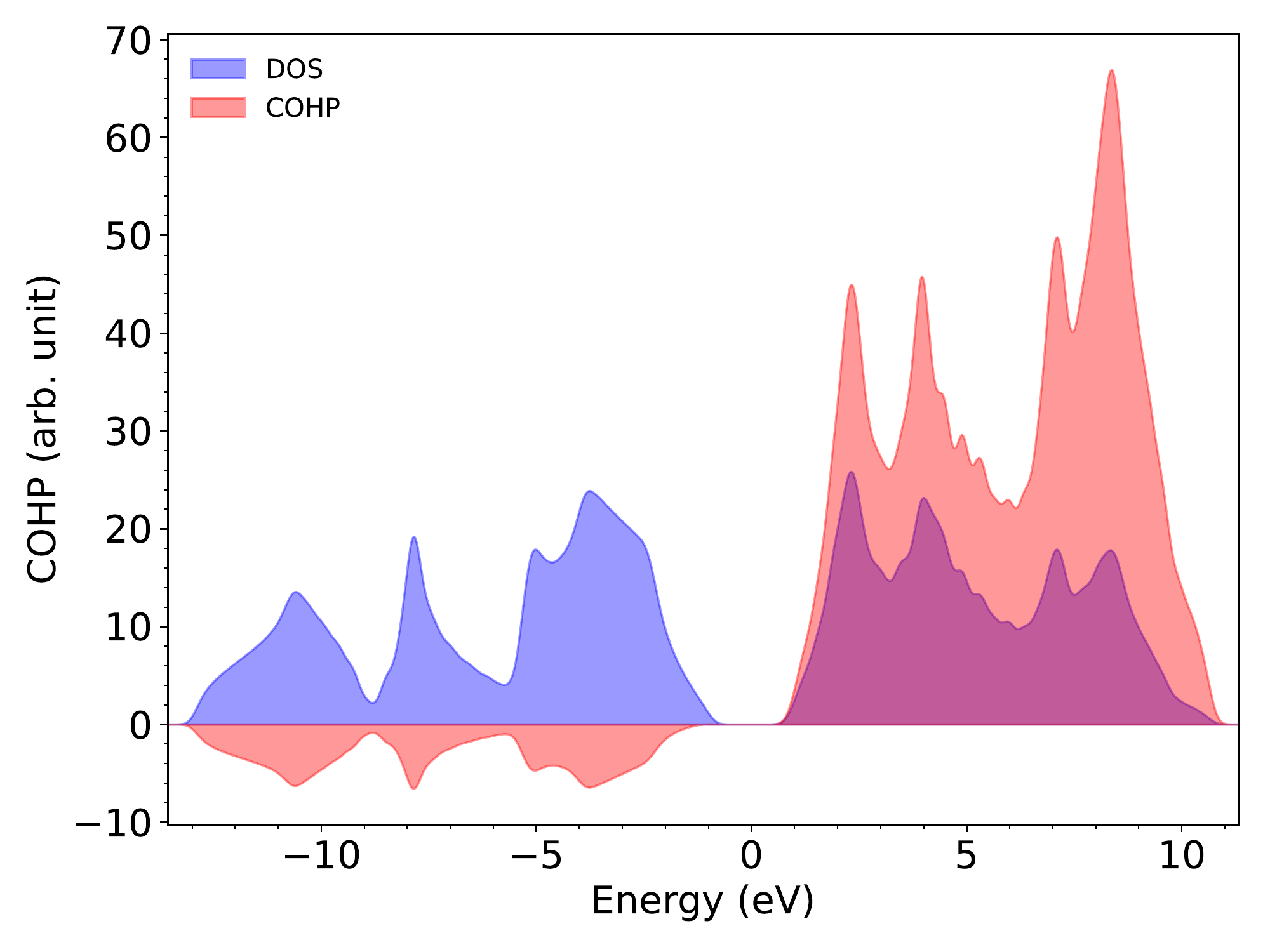}
    \caption{Density of states (DOS) and crystal orbital Hamilton population (COHP) as a function of energy. The negative regime of COHP denotes bonding state, while the positive regime denotes the anti-bonding state.}
    \label{fig:cohp}   
\end{figure}

%\end{document}

%\bibliography{clean}
%merlin.mbs apsrev4-1.bst 2010-07-25 4.21a (PWD, AO, DPC) hacked
%Control: key (0)
%Control: author (8) initials jnrlst
%Control: editor formatted (1) identically to author
%Control: production of article title (-1) disabled
%Control: page (0) single
%Control: year (1) truncated
%Control: production of eprint (0) enabled
\begin{thebibliography}{53}%
	\makeatletter
	\providecommand \@ifxundefined [1]{%
		\@ifx{#1\undefined}
	}%
	\providecommand \@ifnum [1]{%
		\ifnum #1\expandafter \@firstoftwo
		\else \expandafter \@secondoftwo
		\fi
	}%
	\providecommand \@ifx [1]{%
		\ifx #1\expandafter \@firstoftwo
		\else \expandafter \@secondoftwo
		\fi
	}%
	\providecommand \natexlab [1]{#1}%
	\providecommand \enquote  [1]{``#1''}%
	\providecommand \bibnamefont  [1]{#1}%
	\providecommand \bibfnamefont [1]{#1}%
	\providecommand \citenamefont [1]{#1}%
	\providecommand \href@noop [0]{\@secondoftwo}%
	\providecommand \href [0]{\begingroup \@sanitize@url \@href}%
	\providecommand \@href[1]{\@@startlink{#1}\@@href}%
	\providecommand \@@href[1]{\endgroup#1\@@endlink}%
	\providecommand \@sanitize@url [0]{\catcode `\\12\catcode `\$12\catcode
		`\&12\catcode `\#12\catcode `\^12\catcode `\_12\catcode `\%12\relax}%
	\providecommand \@@startlink[1]{}%
	\providecommand \@@endlink[0]{}%
	\providecommand \url  [0]{\begingroup\@sanitize@url \@url }%
	\providecommand \@url [1]{\endgroup\@href {#1}{\urlprefix }}%
	\providecommand \urlprefix  [0]{URL }%
	\providecommand \Eprint [0]{\href }%
	\providecommand \doibase [0]{http://dx.doi.org/}%
	\providecommand \selectlanguage [0]{\@gobble}%
	\providecommand \bibinfo  [0]{\@secondoftwo}%
	\providecommand \bibfield  [0]{\@secondoftwo}%
	\providecommand \translation [1]{[#1]}%
	\providecommand \BibitemOpen [0]{}%
	\providecommand \bibitemStop [0]{}%
	\providecommand \bibitemNoStop [0]{.\EOS\space}%
	\providecommand \EOS [0]{\spacefactor3000\relax}%
	\providecommand \BibitemShut  [1]{\csname bibitem#1\endcsname}%
	\let\auto@bib@innerbib\@empty
	%</preamble>
	\bibitem [{\citenamefont {Ulbricht}\ \emph {et~al.}(2011)\citenamefont
		{Ulbricht}, \citenamefont {Hendry}, \citenamefont {Shan}, \citenamefont
		{Heinz},\ and\ \citenamefont {Bonn}}]{Ulbricht2011}%
	\BibitemOpen
	\bibfield  {author} {\bibinfo {author} {\bibfnamefont {R.}~\bibnamefont
			{Ulbricht}}, \bibinfo {author} {\bibfnamefont {E.}~\bibnamefont {Hendry}},
		\bibinfo {author} {\bibfnamefont {J.}~\bibnamefont {Shan}}, \bibinfo {author}
		{\bibfnamefont {T.~F.}\ \bibnamefont {Heinz}}, \ and\ \bibinfo {author}
		{\bibfnamefont {M.}~\bibnamefont {Bonn}},\ }\href
	{https://doi.org/10.1103%2Frevmodphys.83.543} {\bibfield  {journal} {\bibinfo
		{journal} {Rev. Mod. Phys.}\ }\textbf {\bibinfo {volume} {83}},\ \bibinfo
	{pages} {543} (\bibinfo {year} {2011})}\BibitemShut {NoStop}%
\bibitem [{\citenamefont {Bernardi}\ \emph {et~al.}(2014)\citenamefont
	{Bernardi}, \citenamefont {Vigil-Fowler}, \citenamefont {Lischner},
	\citenamefont {Neaton},\ and\ \citenamefont {Louie}}]{Bernardi2014}%
\BibitemOpen
\bibfield  {author} {\bibinfo {author} {\bibfnamefont {M.}~\bibnamefont
		{Bernardi}}, \bibinfo {author} {\bibfnamefont {D.}~\bibnamefont
		{Vigil-Fowler}}, \bibinfo {author} {\bibfnamefont {J.}~\bibnamefont
		{Lischner}}, \bibinfo {author} {\bibfnamefont {J.~B.}\ \bibnamefont
		{Neaton}}, \ and\ \bibinfo {author} {\bibfnamefont {S.~G.}\ \bibnamefont
		{Louie}},\ }\href {https://doi.org/10.1103%2Fphysrevlett.112.257402}
	{\bibfield  {journal} {\bibinfo  {journal} {Phys. Rev. Lett.}\ }\textbf
		{\bibinfo {volume} {112}},\ \bibinfo {pages} {257402} (\bibinfo {year}
		{2014})}\BibitemShut {NoStop}%
	\bibitem [{\citenamefont {Jadidi}\ \emph {et~al.}(2016)\citenamefont {Jadidi},
		\citenamefont {Suess}, \citenamefont {Tan}, \citenamefont {Cai},
		\citenamefont {Watanabe}, \citenamefont {Taniguchi}, \citenamefont {Sushkov},
		\citenamefont {Mittendorff}, \citenamefont {Hone}, \citenamefont {Drew},
		\citenamefont {Fuhrer},\ and\ \citenamefont {Murphy}}]{Jadidi2016}%
	\BibitemOpen
	\bibfield  {author} {\bibinfo {author} {\bibfnamefont {M.~M.}\ \bibnamefont
			{Jadidi}}, \bibinfo {author} {\bibfnamefont {R.~J.}\ \bibnamefont {Suess}},
		\bibinfo {author} {\bibfnamefont {C.}~\bibnamefont {Tan}}, \bibinfo {author}
		{\bibfnamefont {X.}~\bibnamefont {Cai}}, \bibinfo {author} {\bibfnamefont
			{K.}~\bibnamefont {Watanabe}}, \bibinfo {author} {\bibfnamefont
			{T.}~\bibnamefont {Taniguchi}}, \bibinfo {author} {\bibfnamefont {A.~B.}\
			\bibnamefont {Sushkov}}, \bibinfo {author} {\bibfnamefont {M.}~\bibnamefont
			{Mittendorff}}, \bibinfo {author} {\bibfnamefont {J.}~\bibnamefont {Hone}},
		\bibinfo {author} {\bibfnamefont {H.~D.}\ \bibnamefont {Drew}}, \bibinfo
		{author} {\bibfnamefont {M.~S.}\ \bibnamefont {Fuhrer}}, \ and\ \bibinfo
		{author} {\bibfnamefont {T.~E.}\ \bibnamefont {Murphy}},\ }\href
	{https://doi.org/10.1103%2Fphysrevlett.117.257401} {\bibfield  {journal}
		{\bibinfo  {journal} {Phys. Rev. Lett.}\ }\textbf {\bibinfo {volume} {117}},\
		\bibinfo {pages} {257401} (\bibinfo {year} {2016})}\BibitemShut {NoStop}%
	\bibitem [{\citenamefont {Brown}\ \emph {et~al.}(2015)\citenamefont {Brown},
		\citenamefont {Sundararaman}, \citenamefont {Narang}, \citenamefont
		{Goddard},\ and\ \citenamefont {Atwater}}]{Brown2016a}%
	\BibitemOpen
	\bibfield  {author} {\bibinfo {author} {\bibfnamefont {A.~M.}\ \bibnamefont
			{Brown}}, \bibinfo {author} {\bibfnamefont {R.}~\bibnamefont {Sundararaman}},
		\bibinfo {author} {\bibfnamefont {P.}~\bibnamefont {Narang}}, \bibinfo
		{author} {\bibfnamefont {W.~A.}\ \bibnamefont {Goddard}}, \ and\ \bibinfo
		{author} {\bibfnamefont {H.~A.}\ \bibnamefont {Atwater}},\ }\href
	{https://doi.org/10.1021%2Facsnano.5b06199} {\bibfield  {journal} {\bibinfo
		{journal} {ACS Nano}\ }\textbf {\bibinfo {volume} {10}},\ \bibinfo {pages}
	{957} (\bibinfo {year} {2015})}\BibitemShut {NoStop}%
\bibitem [{\citenamefont {Hwang}\ and\ \citenamefont
	{Sarma}(2007)}]{Hwang2007}%
\BibitemOpen
\bibfield  {author} {\bibinfo {author} {\bibfnamefont {E.~H.}\ \bibnamefont
		{Hwang}}\ and\ \bibinfo {author} {\bibfnamefont {S.~D.}\ \bibnamefont
		{Sarma}},\ }\href {https://doi.org/10.1103%2Fphysrevb.75.205418} {\bibfield
	{journal} {\bibinfo  {journal} {Phys. Rev. B}\ }\textbf {\bibinfo {volume}
		{75}},\ \bibinfo {pages} {205418} (\bibinfo {year} {2007})}\BibitemShut
{NoStop}%
\bibitem [{\citenamefont {Johannsen}\ \emph {et~al.}(2013)\citenamefont
	{Johannsen}, \citenamefont {Ulstrup}, \citenamefont {Cilento}, \citenamefont
	{Crepaldi}, \citenamefont {Zacchigna}, \citenamefont {Cacho}, \citenamefont
	{Turcu}, \citenamefont {Springate}, \citenamefont {Fromm}, \citenamefont
	{Raidel}, \citenamefont {Seyller}, \citenamefont {Parmigiani}, \citenamefont
	{Grioni},\ and\ \citenamefont {Hofmann}}]{Johannsen2013}%
\BibitemOpen
\bibfield  {author} {\bibinfo {author} {\bibfnamefont {J.~C.}\ \bibnamefont
		{Johannsen}}, \bibinfo {author} {\bibfnamefont {S.}~\bibnamefont {Ulstrup}},
	\bibinfo {author} {\bibfnamefont {F.}~\bibnamefont {Cilento}}, \bibinfo
	{author} {\bibfnamefont {A.}~\bibnamefont {Crepaldi}}, \bibinfo {author}
	{\bibfnamefont {M.}~\bibnamefont {Zacchigna}}, \bibinfo {author}
	{\bibfnamefont {C.}~\bibnamefont {Cacho}}, \bibinfo {author} {\bibfnamefont
		{I.~C.~E.}\ \bibnamefont {Turcu}}, \bibinfo {author} {\bibfnamefont
		{E.}~\bibnamefont {Springate}}, \bibinfo {author} {\bibfnamefont
		{F.}~\bibnamefont {Fromm}}, \bibinfo {author} {\bibfnamefont
		{C.}~\bibnamefont {Raidel}}, \bibinfo {author} {\bibfnamefont
		{T.}~\bibnamefont {Seyller}}, \bibinfo {author} {\bibfnamefont
		{F.}~\bibnamefont {Parmigiani}}, \bibinfo {author} {\bibfnamefont
		{M.}~\bibnamefont {Grioni}}, \ and\ \bibinfo {author} {\bibfnamefont
		{P.}~\bibnamefont {Hofmann}},\ }\href
{https://doi.org/10.1103%2Fphysrevlett.111.027403} {\bibfield  {journal}
	{\bibinfo  {journal} {Phys. Rev. Lett.}\ }\textbf {\bibinfo {volume} {111}},\
	\bibinfo {pages} {027403} (\bibinfo {year} {2013})}\BibitemShut {NoStop}%
\bibitem [{\citenamefont {Brida}\ \emph {et~al.}(2013)\citenamefont {Brida},
	\citenamefont {Tomadin}, \citenamefont {Manzoni}, \citenamefont {Kim},
	\citenamefont {Lombardo}, \citenamefont {Milana}, \citenamefont {Nair},
	\citenamefont {Novoselov}, \citenamefont {Ferrari}, \citenamefont {Cerullo},\
	and\ \citenamefont {Polini}}]{Brida2013}%
\BibitemOpen
\bibfield  {author} {\bibinfo {author} {\bibfnamefont {D.}~\bibnamefont
		{Brida}}, \bibinfo {author} {\bibfnamefont {A.}~\bibnamefont {Tomadin}},
	\bibinfo {author} {\bibfnamefont {C.}~\bibnamefont {Manzoni}}, \bibinfo
	{author} {\bibfnamefont {Y.~J.}\ \bibnamefont {Kim}}, \bibinfo {author}
	{\bibfnamefont {A.}~\bibnamefont {Lombardo}}, \bibinfo {author}
	{\bibfnamefont {S.}~\bibnamefont {Milana}}, \bibinfo {author} {\bibfnamefont
		{R.~R.}\ \bibnamefont {Nair}}, \bibinfo {author} {\bibfnamefont {K.~S.}\
		\bibnamefont {Novoselov}}, \bibinfo {author} {\bibfnamefont {A.~C.}\
		\bibnamefont {Ferrari}}, \bibinfo {author} {\bibfnamefont {G.}~\bibnamefont
		{Cerullo}}, \ and\ \bibinfo {author} {\bibfnamefont {M.}~\bibnamefont
		{Polini}},\ }\href {https://doi.org/10.1038%2Fncomms2987} {\bibfield
	{journal} {\bibinfo  {journal} {Nat Commun}\ }\textbf {\bibinfo {volume}
		{4}},\ \bibinfo {pages} {1987} (\bibinfo {year} {2013})}\BibitemShut
{NoStop}%
\bibitem [{\citenamefont {Zeng}\ \emph {et~al.}(2012)\citenamefont {Zeng},
	\citenamefont {Dai}, \citenamefont {Yao}, \citenamefont {Xiao},\ and\
	\citenamefont {Cui}}]{Zeng2012}%
\BibitemOpen
\bibfield  {author} {\bibinfo {author} {\bibfnamefont {H.}~\bibnamefont
		{Zeng}}, \bibinfo {author} {\bibfnamefont {J.}~\bibnamefont {Dai}}, \bibinfo
	{author} {\bibfnamefont {W.}~\bibnamefont {Yao}}, \bibinfo {author}
	{\bibfnamefont {D.}~\bibnamefont {Xiao}}, \ and\ \bibinfo {author}
	{\bibfnamefont {X.}~\bibnamefont {Cui}},\ }\href
{https://doi.org/10.1038%2Fnnano.2012.95} {\bibfield  {journal} {\bibinfo
	{journal} {Nature Nanotech}\ }\textbf {\bibinfo {volume} {7}},\ \bibinfo
{pages} {490} (\bibinfo {year} {2012})}\BibitemShut {NoStop}%
\bibitem [{\citenamefont {Bertoni}\ \emph {et~al.}(2016)\citenamefont
	{Bertoni}, \citenamefont {Nicholson}, \citenamefont {Waldecker},
	\citenamefont {Hübener}, \citenamefont {Monney}, \citenamefont {Giovannini},
	\citenamefont {Puppin}, \citenamefont {Hoesch}, \citenamefont {Springate},
	\citenamefont {Chapman}, \citenamefont {Cacho}, \citenamefont {Wolf},
	\citenamefont {Rubio},\ and\ \citenamefont {Ernstorfer}}]{Bertoni2016}%
\BibitemOpen
\bibfield  {author} {\bibinfo {author} {\bibfnamefont {R.}~\bibnamefont
		{Bertoni}}, \bibinfo {author} {\bibfnamefont {C.}~\bibnamefont {Nicholson}},
	\bibinfo {author} {\bibfnamefont {L.}~\bibnamefont {Waldecker}}, \bibinfo
	{author} {\bibfnamefont {H.}~\bibnamefont {Hübener}}, \bibinfo {author}
	{\bibfnamefont {C.}~\bibnamefont {Monney}}, \bibinfo {author} {\bibfnamefont
		{U.~D.}\ \bibnamefont {Giovannini}}, \bibinfo {author} {\bibfnamefont
		{M.}~\bibnamefont {Puppin}}, \bibinfo {author} {\bibfnamefont
		{M.}~\bibnamefont {Hoesch}}, \bibinfo {author} {\bibfnamefont
		{E.}~\bibnamefont {Springate}}, \bibinfo {author} {\bibfnamefont
		{R.}~\bibnamefont {Chapman}}, \bibinfo {author} {\bibfnamefont
		{C.}~\bibnamefont {Cacho}}, \bibinfo {author} {\bibfnamefont
		{M.}~\bibnamefont {Wolf}}, \bibinfo {author} {\bibfnamefont {A.}~\bibnamefont
		{Rubio}}, \ and\ \bibinfo {author} {\bibfnamefont {R.}~\bibnamefont
		{Ernstorfer}},\ }\href {https://doi.org/10.1103%2Fphysrevlett.117.277201}
	{\bibfield  {journal} {\bibinfo  {journal} {Phys. Rev. Lett.}\ }\textbf
		{\bibinfo {volume} {117}},\ \bibinfo {pages} {277201} (\bibinfo {year}
		{2016})}\BibitemShut {NoStop}%
	\bibitem [{\citenamefont {Mathias}\ \emph {et~al.}(2016)\citenamefont
		{Mathias}, \citenamefont {Eich}, \citenamefont {Urbancic}, \citenamefont
		{Michael}, \citenamefont {Carr}, \citenamefont {Emmerich}, \citenamefont
		{Stange}, \citenamefont {Popmintchev}, \citenamefont {Rohwer}, \citenamefont
		{Wiesenmayer}, \citenamefont {Ruffing}, \citenamefont {Jakobs}, \citenamefont
		{Hellmann}, \citenamefont {Matyba}, \citenamefont {Chen}, \citenamefont
		{Kipp}, \citenamefont {Bauer}, \citenamefont {Kapteyn}, \citenamefont
		{Schneider}, \citenamefont {Rossnagel}, \citenamefont {Murnane},\ and\
		\citenamefont {Aeschlimann}}]{Mathias2016}%
	\BibitemOpen
	\bibfield  {author} {\bibinfo {author} {\bibfnamefont {S.}~\bibnamefont
			{Mathias}}, \bibinfo {author} {\bibfnamefont {S.}~\bibnamefont {Eich}},
		\bibinfo {author} {\bibfnamefont {J.}~\bibnamefont {Urbancic}}, \bibinfo
		{author} {\bibfnamefont {S.}~\bibnamefont {Michael}}, \bibinfo {author}
		{\bibfnamefont {A.~V.}\ \bibnamefont {Carr}}, \bibinfo {author}
		{\bibfnamefont {S.}~\bibnamefont {Emmerich}}, \bibinfo {author}
		{\bibfnamefont {A.}~\bibnamefont {Stange}}, \bibinfo {author} {\bibfnamefont
			{T.}~\bibnamefont {Popmintchev}}, \bibinfo {author} {\bibfnamefont
			{T.}~\bibnamefont {Rohwer}}, \bibinfo {author} {\bibfnamefont
			{M.}~\bibnamefont {Wiesenmayer}}, \bibinfo {author} {\bibfnamefont
			{A.}~\bibnamefont {Ruffing}}, \bibinfo {author} {\bibfnamefont
			{S.}~\bibnamefont {Jakobs}}, \bibinfo {author} {\bibfnamefont
			{S.}~\bibnamefont {Hellmann}}, \bibinfo {author} {\bibfnamefont
			{P.}~\bibnamefont {Matyba}}, \bibinfo {author} {\bibfnamefont
			{C.}~\bibnamefont {Chen}}, \bibinfo {author} {\bibfnamefont {L.}~\bibnamefont
			{Kipp}}, \bibinfo {author} {\bibfnamefont {M.}~\bibnamefont {Bauer}},
		\bibinfo {author} {\bibfnamefont {H.~C.}\ \bibnamefont {Kapteyn}}, \bibinfo
		{author} {\bibfnamefont {H.~C.}\ \bibnamefont {Schneider}}, \bibinfo {author}
		{\bibfnamefont {K.}~\bibnamefont {Rossnagel}}, \bibinfo {author}
		{\bibfnamefont {M.~M.}\ \bibnamefont {Murnane}}, \ and\ \bibinfo {author}
		{\bibfnamefont {M.}~\bibnamefont {Aeschlimann}},\ }\href
	{https://doi.org/10.1038%2Fncomms12902} {\bibfield  {journal} {\bibinfo
		{journal} {Nat Comms}\ }\textbf {\bibinfo {volume} {7}},\ \bibinfo {pages}
	{12902} (\bibinfo {year} {2016})}\BibitemShut {NoStop}%
\bibitem [{\citenamefont {Hao}\ \emph {et~al.}(2016)\citenamefont {Hao},
	\citenamefont {Moody}, \citenamefont {Wu}, \citenamefont {Dass},
	\citenamefont {Xu}, \citenamefont {Chen}, \citenamefont {Sun}, \citenamefont
	{Li}, \citenamefont {Li}, \citenamefont {MacDonald},\ and\ \citenamefont
	{Li}}]{Hao2016}%
\BibitemOpen
\bibfield  {author} {\bibinfo {author} {\bibfnamefont {K.}~\bibnamefont
		{Hao}}, \bibinfo {author} {\bibfnamefont {G.}~\bibnamefont {Moody}}, \bibinfo
	{author} {\bibfnamefont {F.}~\bibnamefont {Wu}}, \bibinfo {author}
	{\bibfnamefont {C.~K.}\ \bibnamefont {Dass}}, \bibinfo {author}
	{\bibfnamefont {L.}~\bibnamefont {Xu}}, \bibinfo {author} {\bibfnamefont
		{C.-H.}\ \bibnamefont {Chen}}, \bibinfo {author} {\bibfnamefont
		{L.}~\bibnamefont {Sun}}, \bibinfo {author} {\bibfnamefont {M.-Y.}\
		\bibnamefont {Li}}, \bibinfo {author} {\bibfnamefont {L.-J.}\ \bibnamefont
		{Li}}, \bibinfo {author} {\bibfnamefont {A.~H.}\ \bibnamefont {MacDonald}}, \
	and\ \bibinfo {author} {\bibfnamefont {X.}~\bibnamefont {Li}},\ }\href
{https://doi.org/10.1038%2Fnphys3674} {\bibfield  {journal} {\bibinfo
	{journal} {Nature Phys}\ }\textbf {\bibinfo {volume} {12}},\ \bibinfo {pages}
{677} (\bibinfo {year} {2016})}\BibitemShut {NoStop}%
\bibitem [{\citenamefont {Zhang}\ \emph {et~al.}(2017)\citenamefont {Zhang},
	\citenamefont {Hong}, \citenamefont {Lian}, \citenamefont {Ma}, \citenamefont
	{Xu}, \citenamefont {Zhou}, \citenamefont {Fu}, \citenamefont {Liu},\ and\
	\citenamefont {Meng}}]{Zhang2017a}%
\BibitemOpen
\bibfield  {author} {\bibinfo {author} {\bibfnamefont {J.}~\bibnamefont
		{Zhang}}, \bibinfo {author} {\bibfnamefont {H.}~\bibnamefont {Hong}},
	\bibinfo {author} {\bibfnamefont {C.}~\bibnamefont {Lian}}, \bibinfo {author}
	{\bibfnamefont {W.}~\bibnamefont {Ma}}, \bibinfo {author} {\bibfnamefont
		{X.}~\bibnamefont {Xu}}, \bibinfo {author} {\bibfnamefont {X.}~\bibnamefont
		{Zhou}}, \bibinfo {author} {\bibfnamefont {H.}~\bibnamefont {Fu}}, \bibinfo
	{author} {\bibfnamefont {K.}~\bibnamefont {Liu}}, \ and\ \bibinfo {author}
	{\bibfnamefont {S.}~\bibnamefont {Meng}},\ }\href
{https://doi.org/10.1002%2Fadvs.201700086} {\bibfield  {journal} {\bibinfo
	{journal} {Adv. Sci.}\ }\textbf {\bibinfo {volume} {4}},\ \bibinfo {pages}
{1700086} (\bibinfo {year} {2017})}\BibitemShut {NoStop}%
\bibitem [{\citenamefont {Shank}\ \emph {et~al.}(1983)\citenamefont {Shank},
	\citenamefont {Yen},\ and\ \citenamefont {Hirlimann}}]{Shank1983}%
\BibitemOpen
\bibfield  {author} {\bibinfo {author} {\bibfnamefont {C.~V.}\ \bibnamefont
		{Shank}}, \bibinfo {author} {\bibfnamefont {R.}~\bibnamefont {Yen}}, \ and\
	\bibinfo {author} {\bibfnamefont {C.}~\bibnamefont {Hirlimann}},\ }\href
{https://doi.org/10.1103%2Fphysrevlett.50.454} {\bibfield  {journal}
	{\bibinfo  {journal} {Phys. Rev. Lett.}\ }\textbf {\bibinfo {volume} {50}},\
	\bibinfo {pages} {454} (\bibinfo {year} {1983})}\BibitemShut {NoStop}%
\bibitem [{\citenamefont {Möhr-Vorobeva}\ \emph {et~al.}(2011)\citenamefont
	{Möhr-Vorobeva}, \citenamefont {Johnson}, \citenamefont {Beaud},
	\citenamefont {Staub}, \citenamefont {Souza}, \citenamefont {Milne},
	\citenamefont {Ingold}, \citenamefont {Demsar}, \citenamefont {Schaefer},\
	and\ \citenamefont {Titov}}]{Mohr-Vorobeva2011}%
\BibitemOpen
\bibfield  {author} {\bibinfo {author} {\bibfnamefont {E.}~\bibnamefont
		{Möhr-Vorobeva}}, \bibinfo {author} {\bibfnamefont {S.~L.}\ \bibnamefont
		{Johnson}}, \bibinfo {author} {\bibfnamefont {P.}~\bibnamefont {Beaud}},
	\bibinfo {author} {\bibfnamefont {U.}~\bibnamefont {Staub}}, \bibinfo
	{author} {\bibfnamefont {R.~D.}\ \bibnamefont {Souza}}, \bibinfo {author}
	{\bibfnamefont {C.}~\bibnamefont {Milne}}, \bibinfo {author} {\bibfnamefont
		{G.}~\bibnamefont {Ingold}}, \bibinfo {author} {\bibfnamefont
		{J.}~\bibnamefont {Demsar}}, \bibinfo {author} {\bibfnamefont
		{H.}~\bibnamefont {Schaefer}}, \ and\ \bibinfo {author} {\bibfnamefont
		{A.}~\bibnamefont {Titov}},\ }\href
{https://doi.org/10.1103%2Fphysrevlett.107.036403} {\bibfield  {journal}
	{\bibinfo  {journal} {Phys. Rev. Lett.}\ }\textbf {\bibinfo {volume} {107}},\
	\bibinfo {pages} {036403} (\bibinfo {year} {2011})}\BibitemShut {NoStop}%
\bibitem [{\citenamefont {Zalden}\ \emph {et~al.}(2015)\citenamefont {Zalden},
	\citenamefont {von Hoegen}, \citenamefont {Landreman}, \citenamefont
	{Wuttig},\ and\ \citenamefont {Lindenberg}}]{Zalden2015}%
\BibitemOpen
\bibfield  {author} {\bibinfo {author} {\bibfnamefont {P.}~\bibnamefont
		{Zalden}}, \bibinfo {author} {\bibfnamefont {A.}~\bibnamefont {von Hoegen}},
	\bibinfo {author} {\bibfnamefont {P.}~\bibnamefont {Landreman}}, \bibinfo
	{author} {\bibfnamefont {M.}~\bibnamefont {Wuttig}}, \ and\ \bibinfo {author}
	{\bibfnamefont {A.~M.}\ \bibnamefont {Lindenberg}},\ }\href
{https://doi.org/10.1021%2Facs.chemmater.5b02011} {\bibfield  {journal}
	{\bibinfo  {journal} {Chem. Mater.}\ }\textbf {\bibinfo {volume} {27}},\
	\bibinfo {pages} {5641} (\bibinfo {year} {2015})}\BibitemShut {NoStop}%
\bibitem [{\citenamefont {Matsubara}\ \emph
	{et~al.}(2016{\natexlab{a}})\citenamefont {Matsubara}, \citenamefont {Okada},
	\citenamefont {Ichitsubo}, \citenamefont {Kawaguchi}, \citenamefont {Hirata},
	\citenamefont {Guan}, \citenamefont {Tokuda}, \citenamefont {Tanimura},
	\citenamefont {Matsunaga}, \citenamefont {Chen},\ and\ \citenamefont
	{Yamada}}]{Matsubara2016}%
\BibitemOpen
\bibfield  {author} {\bibinfo {author} {\bibfnamefont {E.}~\bibnamefont
		{Matsubara}}, \bibinfo {author} {\bibfnamefont {S.}~\bibnamefont {Okada}},
	\bibinfo {author} {\bibfnamefont {T.}~\bibnamefont {Ichitsubo}}, \bibinfo
	{author} {\bibfnamefont {T.}~\bibnamefont {Kawaguchi}}, \bibinfo {author}
	{\bibfnamefont {A.}~\bibnamefont {Hirata}}, \bibinfo {author} {\bibfnamefont
		{P.}~\bibnamefont {Guan}}, \bibinfo {author} {\bibfnamefont {K.}~\bibnamefont
		{Tokuda}}, \bibinfo {author} {\bibfnamefont {K.}~\bibnamefont {Tanimura}},
	\bibinfo {author} {\bibfnamefont {T.}~\bibnamefont {Matsunaga}}, \bibinfo
	{author} {\bibfnamefont {M.}~\bibnamefont {Chen}}, \ and\ \bibinfo {author}
	{\bibfnamefont {N.}~\bibnamefont {Yamada}},\ }\href
{https://doi.org/10.1103%2Fphysrevlett.117.135501} {\bibfield  {journal}
	{\bibinfo  {journal} {Phys. Rev. Lett.}\ }\textbf {\bibinfo {volume} {117}},\
	\bibinfo {pages} {135501} (\bibinfo {year} {2016}{\natexlab{a}})}\BibitemShut
{NoStop}%
\bibitem [{\citenamefont {Zalden}\ \emph {et~al.}(2016)\citenamefont {Zalden},
	\citenamefont {Shu}, \citenamefont {Chen}, \citenamefont {Wu}, \citenamefont
	{Zhu}, \citenamefont {Wen}, \citenamefont {Johnston}, \citenamefont {Shen},
	\citenamefont {Landreman}, \citenamefont {Brongersma}, \citenamefont {Fong},
	\citenamefont {Wong}, \citenamefont {Sher}, \citenamefont {Jost},
	\citenamefont {Kaes}, \citenamefont {Salinga}, \citenamefont {von Hoegen},
	\citenamefont {Wuttig},\ and\ \citenamefont {Lindenberg}}]{Zalden2016}%
\BibitemOpen
\bibfield  {author} {\bibinfo {author} {\bibfnamefont {P.}~\bibnamefont
		{Zalden}}, \bibinfo {author} {\bibfnamefont {M.~J.}\ \bibnamefont {Shu}},
	\bibinfo {author} {\bibfnamefont {F.}~\bibnamefont {Chen}}, \bibinfo {author}
	{\bibfnamefont {X.}~\bibnamefont {Wu}}, \bibinfo {author} {\bibfnamefont
		{Y.}~\bibnamefont {Zhu}}, \bibinfo {author} {\bibfnamefont {H.}~\bibnamefont
		{Wen}}, \bibinfo {author} {\bibfnamefont {S.}~\bibnamefont {Johnston}},
	\bibinfo {author} {\bibfnamefont {Z.-X.}\ \bibnamefont {Shen}}, \bibinfo
	{author} {\bibfnamefont {P.}~\bibnamefont {Landreman}}, \bibinfo {author}
	{\bibfnamefont {M.}~\bibnamefont {Brongersma}}, \bibinfo {author}
	{\bibfnamefont {S.~W.}\ \bibnamefont {Fong}}, \bibinfo {author}
	{\bibfnamefont {H.-S.~P.}\ \bibnamefont {Wong}}, \bibinfo {author}
	{\bibfnamefont {M.-J.}\ \bibnamefont {Sher}}, \bibinfo {author}
	{\bibfnamefont {P.}~\bibnamefont {Jost}}, \bibinfo {author} {\bibfnamefont
		{M.}~\bibnamefont {Kaes}}, \bibinfo {author} {\bibfnamefont {M.}~\bibnamefont
		{Salinga}}, \bibinfo {author} {\bibfnamefont {A.}~\bibnamefont {von Hoegen}},
	\bibinfo {author} {\bibfnamefont {M.}~\bibnamefont {Wuttig}}, \ and\ \bibinfo
	{author} {\bibfnamefont {A.~M.}\ \bibnamefont {Lindenberg}},\ }\href
{https://doi.org/10.1103%2Fphysrevlett.117.067601} {\bibfield  {journal}
	{\bibinfo  {journal} {Phys. Rev. Lett.}\ }\textbf {\bibinfo {volume} {117}},\
	\bibinfo {pages} {067601} (\bibinfo {year} {2016})}\BibitemShut {NoStop}%
\bibitem [{\citenamefont {Bang}\ \emph {et~al.}(2016)\citenamefont {Bang},
	\citenamefont {Sun}, \citenamefont {Liu}, \citenamefont {Gao},\ and\
	\citenamefont {Zhang}}]{Bang2016}%
\BibitemOpen
\bibfield  {author} {\bibinfo {author} {\bibfnamefont {J.}~\bibnamefont
		{Bang}}, \bibinfo {author} {\bibfnamefont {Y.}~\bibnamefont {Sun}}, \bibinfo
	{author} {\bibfnamefont {X.-Q.}\ \bibnamefont {Liu}}, \bibinfo {author}
	{\bibfnamefont {F.}~\bibnamefont {Gao}}, \ and\ \bibinfo {author}
	{\bibfnamefont {S.}~\bibnamefont {Zhang}},\ }\href
{https://doi.org/10.1103%2Fphysrevlett.117.126402} {\bibfield  {journal}
	{\bibinfo  {journal} {Phys. Rev. Lett.}\ }\textbf {\bibinfo {volume} {117}},\
	\bibinfo {pages} {126402} (\bibinfo {year} {2016})}\BibitemShut {NoStop}%
\bibitem [{\citenamefont {Chen}\ \emph
	{et~al.}(2018{\natexlab{a}})\citenamefont {Chen}, \citenamefont {Li},
	\citenamefont {Bang}, \citenamefont {Wang}, \citenamefont {Han},
	\citenamefont {West}, \citenamefont {Zhang},\ and\ \citenamefont
	{Sun}}]{Chen2018}%
\BibitemOpen
\bibfield  {author} {\bibinfo {author} {\bibfnamefont {N.-K.}\ \bibnamefont
		{Chen}}, \bibinfo {author} {\bibfnamefont {X.-B.}\ \bibnamefont {Li}},
	\bibinfo {author} {\bibfnamefont {J.}~\bibnamefont {Bang}}, \bibinfo {author}
	{\bibfnamefont {X.-P.}\ \bibnamefont {Wang}}, \bibinfo {author}
	{\bibfnamefont {D.}~\bibnamefont {Han}}, \bibinfo {author} {\bibfnamefont
		{D.}~\bibnamefont {West}}, \bibinfo {author} {\bibfnamefont {S.}~\bibnamefont
		{Zhang}}, \ and\ \bibinfo {author} {\bibfnamefont {H.-B.}\ \bibnamefont
		{Sun}},\ }\href {https://doi.org/10.1103%2Fphysrevlett.120.185701} {\bibfield
	{journal} {\bibinfo  {journal} {Phys. Rev. Lett.}\ }\textbf {\bibinfo
		{volume} {120}},\ \bibinfo {pages} {185701} (\bibinfo {year}
	{2018}{\natexlab{a}})}\BibitemShut {NoStop}%
\bibitem [{\citenamefont {Qi}\ \emph {et~al.}(2009)\citenamefont {Qi},
	\citenamefont {Shin}, \citenamefont {Yeh}, \citenamefont {Nelson},\ and\
	\citenamefont {Rappe}}]{Qi2009a}%
\BibitemOpen
\bibfield  {author} {\bibinfo {author} {\bibfnamefont {T.}~\bibnamefont
		{Qi}}, \bibinfo {author} {\bibfnamefont {Y.-H.}\ \bibnamefont {Shin}},
	\bibinfo {author} {\bibfnamefont {K.-L.}\ \bibnamefont {Yeh}}, \bibinfo
	{author} {\bibfnamefont {K.~A.}\ \bibnamefont {Nelson}}, \ and\ \bibinfo
	{author} {\bibfnamefont {A.~M.}\ \bibnamefont {Rappe}},\ }\href
{https://doi.org/10.1103%2Fphysrevlett.102.247603} {\bibfield  {journal}
	{\bibinfo  {journal} {Phys. Rev. Lett.}\ }\textbf {\bibinfo {volume} {102}},\
	\bibinfo {pages} {247603} (\bibinfo {year} {2009})}\BibitemShut {NoStop}%
\bibitem [{\citenamefont {Rapp}\ \emph {et~al.}(2015)\citenamefont {Rapp},
	\citenamefont {Haberl}, \citenamefont {Pickard}, \citenamefont {Bradby},
	\citenamefont {Gamaly}, \citenamefont {Williams},\ and\ \citenamefont
	{Rode}}]{Rapp2015}%
\BibitemOpen
\bibfield  {author} {\bibinfo {author} {\bibfnamefont {L.}~\bibnamefont
		{Rapp}}, \bibinfo {author} {\bibfnamefont {B.}~\bibnamefont {Haberl}},
	\bibinfo {author} {\bibfnamefont {C.}~\bibnamefont {Pickard}}, \bibinfo
	{author} {\bibfnamefont {J.}~\bibnamefont {Bradby}}, \bibinfo {author}
	{\bibfnamefont {E.}~\bibnamefont {Gamaly}}, \bibinfo {author} {\bibfnamefont
		{J.}~\bibnamefont {Williams}}, \ and\ \bibinfo {author} {\bibfnamefont
		{A.}~\bibnamefont {Rode}},\ }\href {https://doi.org/10.1038%2Fncomms8555}
	{\bibfield  {journal} {\bibinfo  {journal} {Nat Commun}\ }\textbf {\bibinfo
			{volume} {6}},\ \bibinfo {pages} {7555} (\bibinfo {year} {2015})}\BibitemShut
	{NoStop}%
	\bibitem [{\citenamefont {Iwano}\ \emph {et~al.}(2017)\citenamefont {Iwano},
		\citenamefont {Shimoi}, \citenamefont {Miyamoto}, \citenamefont {Hata},
		\citenamefont {Sotome}, \citenamefont {Kida}, \citenamefont {Horiuchi},\ and\
		\citenamefont {Okamoto}}]{Iwano2017}%
	\BibitemOpen
	\bibfield  {author} {\bibinfo {author} {\bibfnamefont {K.}~\bibnamefont
			{Iwano}}, \bibinfo {author} {\bibfnamefont {Y.}~\bibnamefont {Shimoi}},
		\bibinfo {author} {\bibfnamefont {T.}~\bibnamefont {Miyamoto}}, \bibinfo
		{author} {\bibfnamefont {D.}~\bibnamefont {Hata}}, \bibinfo {author}
		{\bibfnamefont {M.}~\bibnamefont {Sotome}}, \bibinfo {author} {\bibfnamefont
			{N.}~\bibnamefont {Kida}}, \bibinfo {author} {\bibfnamefont {S.}~\bibnamefont
			{Horiuchi}}, \ and\ \bibinfo {author} {\bibfnamefont {H.}~\bibnamefont
			{Okamoto}},\ }\href {https://doi.org/10.1103%2Fphysrevlett.118.107404}
		{\bibfield  {journal} {\bibinfo  {journal} {Phys. Rev. Lett.}\ }\textbf
			{\bibinfo {volume} {118}},\ \bibinfo {pages} {107404} (\bibinfo {year}
			{2017})}\BibitemShut {NoStop}%
		\bibitem [{\citenamefont {Porer}\ \emph {et~al.}(2018)\citenamefont {Porer},
			\citenamefont {Fechner}, \citenamefont {Bothschafter}, \citenamefont
			{Rettig}, \citenamefont {Savoini}, \citenamefont {Esposito}, \citenamefont
			{Rittmann}, \citenamefont {Kubli}, \citenamefont {Neugebauer}, \citenamefont
			{Abreu}, \citenamefont {Kubacka}, \citenamefont {Huber}, \citenamefont
			{Lantz}, \citenamefont {Parchenko}, \citenamefont {Grübel}, \citenamefont
			{Paarmann}, \citenamefont {Noack}, \citenamefont {Beaud}, \citenamefont
			{Ingold}, \citenamefont {Aschauer}, \citenamefont {Johnson},\ and\
			\citenamefont {Staub}}]{Porer2018}%
		\BibitemOpen
		\bibfield  {author} {\bibinfo {author} {\bibfnamefont {M.}~\bibnamefont
				{Porer}}, \bibinfo {author} {\bibfnamefont {M.}~\bibnamefont {Fechner}},
			\bibinfo {author} {\bibfnamefont {E.}~\bibnamefont {Bothschafter}}, \bibinfo
			{author} {\bibfnamefont {L.}~\bibnamefont {Rettig}}, \bibinfo {author}
			{\bibfnamefont {M.}~\bibnamefont {Savoini}}, \bibinfo {author} {\bibfnamefont
				{V.}~\bibnamefont {Esposito}}, \bibinfo {author} {\bibfnamefont
				{J.}~\bibnamefont {Rittmann}}, \bibinfo {author} {\bibfnamefont
				{M.}~\bibnamefont {Kubli}}, \bibinfo {author} {\bibfnamefont
				{M.}~\bibnamefont {Neugebauer}}, \bibinfo {author} {\bibfnamefont
				{E.}~\bibnamefont {Abreu}}, \bibinfo {author} {\bibfnamefont
				{T.}~\bibnamefont {Kubacka}}, \bibinfo {author} {\bibfnamefont
				{T.}~\bibnamefont {Huber}}, \bibinfo {author} {\bibfnamefont
				{G.}~\bibnamefont {Lantz}}, \bibinfo {author} {\bibfnamefont
				{S.}~\bibnamefont {Parchenko}}, \bibinfo {author} {\bibfnamefont
				{S.}~\bibnamefont {Grübel}}, \bibinfo {author} {\bibfnamefont
				{A.}~\bibnamefont {Paarmann}}, \bibinfo {author} {\bibfnamefont
				{J.}~\bibnamefont {Noack}}, \bibinfo {author} {\bibfnamefont
				{P.}~\bibnamefont {Beaud}}, \bibinfo {author} {\bibfnamefont
				{G.}~\bibnamefont {Ingold}}, \bibinfo {author} {\bibfnamefont
				{U.}~\bibnamefont {Aschauer}}, \bibinfo {author} {\bibfnamefont
				{S.}~\bibnamefont {Johnson}}, \ and\ \bibinfo {author} {\bibfnamefont
				{U.}~\bibnamefont {Staub}},\ }\href
		{https://doi.org/10.1103%2Fphysrevlett.121.055701} {\bibfield  {journal}
			{\bibinfo  {journal} {Phys. Rev. Lett.}\ }\textbf {\bibinfo {volume} {121}},\
			\bibinfo {pages} {055701} (\bibinfo {year} {2018})}\BibitemShut {NoStop}%
		\bibitem [{\citenamefont {Mankowsky}\ \emph {et~al.}(2017)\citenamefont
			{Mankowsky}, \citenamefont {von Hoegen}, \citenamefont {Först},\ and\
			\citenamefont {Cavalleri}}]{Mankowsky2017}%
		\BibitemOpen
		\bibfield  {author} {\bibinfo {author} {\bibfnamefont {R.}~\bibnamefont
				{Mankowsky}}, \bibinfo {author} {\bibfnamefont {A.}~\bibnamefont {von
					Hoegen}}, \bibinfo {author} {\bibfnamefont {M.}~\bibnamefont {Först}}, \
			and\ \bibinfo {author} {\bibfnamefont {A.}~\bibnamefont {Cavalleri}},\ }\href
		{https://doi.org/10.1103%2Fphysrevlett.118.197601} {\bibfield  {journal}
			{\bibinfo  {journal} {Phys. Rev. Lett.}\ }\textbf {\bibinfo {volume} {118}},\
			\bibinfo {pages} {197601} (\bibinfo {year} {2017})}\BibitemShut {NoStop}%
		\bibitem [{\citenamefont {Matsubara}\ \emph
			{et~al.}(2016{\natexlab{b}})\citenamefont {Matsubara}, \citenamefont {Okada},
			\citenamefont {Ichitsubo}, \citenamefont {Kawaguchi}, \citenamefont {Hirata},
			\citenamefont {Guan}, \citenamefont {Tokuda}, \citenamefont {Tanimura},
			\citenamefont {Matsunaga}, \citenamefont {Chen},\ and\ \citenamefont
			{Yamada}}]{PhysRevLett.117.135501}%
		\BibitemOpen
		\bibfield  {author} {\bibinfo {author} {\bibfnamefont {E.}~\bibnamefont
				{Matsubara}}, \bibinfo {author} {\bibfnamefont {S.}~\bibnamefont {Okada}},
			\bibinfo {author} {\bibfnamefont {T.}~\bibnamefont {Ichitsubo}}, \bibinfo
			{author} {\bibfnamefont {T.}~\bibnamefont {Kawaguchi}}, \bibinfo {author}
			{\bibfnamefont {A.}~\bibnamefont {Hirata}}, \bibinfo {author} {\bibfnamefont
				{P.}~\bibnamefont {Guan}}, \bibinfo {author} {\bibfnamefont {K.}~\bibnamefont
				{Tokuda}}, \bibinfo {author} {\bibfnamefont {K.}~\bibnamefont {Tanimura}},
			\bibinfo {author} {\bibfnamefont {T.}~\bibnamefont {Matsunaga}}, \bibinfo
			{author} {\bibfnamefont {M.}~\bibnamefont {Chen}}, \ and\ \bibinfo {author}
			{\bibfnamefont {N.}~\bibnamefont {Yamada}},\ }\href
		{https://doi.org/10.1103%2Fphysrevlett.117.135501} {\bibfield  {journal}
			{\bibinfo  {journal} {Phys. Rev. Lett.}\ }\textbf {\bibinfo {volume} {117}},\
			\bibinfo {pages} {135501} (\bibinfo {year} {2016}{\natexlab{b}})}\BibitemShut
		{NoStop}%
		\bibitem [{\citenamefont {Chen}\ \emph
			{et~al.}(2018{\natexlab{b}})\citenamefont {Chen}, \citenamefont {Li},
			\citenamefont {Bang}, \citenamefont {Wang}, \citenamefont {Han},
			\citenamefont {West}, \citenamefont {Zhang},\ and\ \citenamefont
			{Sun}}]{PhysRevLett.120.185701}%
		\BibitemOpen
		\bibfield  {author} {\bibinfo {author} {\bibfnamefont {N.-K.}\ \bibnamefont
				{Chen}}, \bibinfo {author} {\bibfnamefont {X.-B.}\ \bibnamefont {Li}},
			\bibinfo {author} {\bibfnamefont {J.}~\bibnamefont {Bang}}, \bibinfo {author}
			{\bibfnamefont {X.-P.}\ \bibnamefont {Wang}}, \bibinfo {author}
			{\bibfnamefont {D.}~\bibnamefont {Han}}, \bibinfo {author} {\bibfnamefont
				{D.}~\bibnamefont {West}}, \bibinfo {author} {\bibfnamefont {S.}~\bibnamefont
				{Zhang}}, \ and\ \bibinfo {author} {\bibfnamefont {H.-B.}\ \bibnamefont
				{Sun}},\ }\href {https://doi.org/10.1103%2Fphysrevlett.120.185701} {\bibfield
			{journal} {\bibinfo  {journal} {Phys. Rev. Lett.}\ }\textbf {\bibinfo
				{volume} {120}},\ \bibinfo {pages} {185701} (\bibinfo {year}
			{2018}{\natexlab{b}})}\BibitemShut {NoStop}%
		\bibitem [{\citenamefont {Lindenberg}(2005)}]{Lindenberg2005}%
		\BibitemOpen
		\bibfield  {author} {\bibinfo {author} {\bibfnamefont {A.~M.}\ \bibnamefont
				{Lindenberg}},\ }\href {https://doi.org/10.1126%2Fscience.1107996} {\bibfield
			{journal} {\bibinfo  {journal} {Science}\ }\textbf {\bibinfo {volume}
				{308}},\ \bibinfo {pages} {392} (\bibinfo {year} {2005})}\BibitemShut
		{NoStop}%
		\bibitem [{\citenamefont {Hillyard}\ \emph {et~al.}(2007)\citenamefont
			{Hillyard}, \citenamefont {Gaffney}, \citenamefont {Lindenberg},
			\citenamefont {Engemann}, \citenamefont {Akre}, \citenamefont {Arthur},
			\citenamefont {Blome}, \citenamefont {Bucksbaum}, \citenamefont {Cavalieri},
			\citenamefont {Deb}, \citenamefont {Falcone}, \citenamefont {Fritz},
			\citenamefont {Fuoss}, \citenamefont {Hajdu}, \citenamefont {Krejcik},
			\citenamefont {Larsson}, \citenamefont {Lee}, \citenamefont {Meyer},
			\citenamefont {Nelson}, \citenamefont {Pahl}, \citenamefont {Reis},
			\citenamefont {Rudati}, \citenamefont {Siddons}, \citenamefont
			{Sokolowski-Tinten}, \citenamefont {von~der Linde},\ and\ \citenamefont
			{Hastings}}]{Hillyard2007}%
		\BibitemOpen
		\bibfield  {author} {\bibinfo {author} {\bibfnamefont {P.~B.}\ \bibnamefont
				{Hillyard}}, \bibinfo {author} {\bibfnamefont {K.~J.}\ \bibnamefont
				{Gaffney}}, \bibinfo {author} {\bibfnamefont {A.~M.}\ \bibnamefont
				{Lindenberg}}, \bibinfo {author} {\bibfnamefont {S.}~\bibnamefont
				{Engemann}}, \bibinfo {author} {\bibfnamefont {R.~A.}\ \bibnamefont {Akre}},
			\bibinfo {author} {\bibfnamefont {J.}~\bibnamefont {Arthur}}, \bibinfo
			{author} {\bibfnamefont {C.}~\bibnamefont {Blome}}, \bibinfo {author}
			{\bibfnamefont {P.~H.}\ \bibnamefont {Bucksbaum}}, \bibinfo {author}
			{\bibfnamefont {A.~L.}\ \bibnamefont {Cavalieri}}, \bibinfo {author}
			{\bibfnamefont {A.}~\bibnamefont {Deb}}, \bibinfo {author} {\bibfnamefont
				{R.~W.}\ \bibnamefont {Falcone}}, \bibinfo {author} {\bibfnamefont {D.~M.}\
				\bibnamefont {Fritz}}, \bibinfo {author} {\bibfnamefont {P.~H.}\ \bibnamefont
				{Fuoss}}, \bibinfo {author} {\bibfnamefont {J.}~\bibnamefont {Hajdu}},
			\bibinfo {author} {\bibfnamefont {P.}~\bibnamefont {Krejcik}}, \bibinfo
			{author} {\bibfnamefont {J.}~\bibnamefont {Larsson}}, \bibinfo {author}
			{\bibfnamefont {S.~H.}\ \bibnamefont {Lee}}, \bibinfo {author} {\bibfnamefont
				{D.~A.}\ \bibnamefont {Meyer}}, \bibinfo {author} {\bibfnamefont {A.~J.}\
				\bibnamefont {Nelson}}, \bibinfo {author} {\bibfnamefont {R.}~\bibnamefont
				{Pahl}}, \bibinfo {author} {\bibfnamefont {D.~A.}\ \bibnamefont {Reis}},
			\bibinfo {author} {\bibfnamefont {J.}~\bibnamefont {Rudati}}, \bibinfo
			{author} {\bibfnamefont {D.~P.}\ \bibnamefont {Siddons}}, \bibinfo {author}
			{\bibfnamefont {K.}~\bibnamefont {Sokolowski-Tinten}}, \bibinfo {author}
			{\bibfnamefont {D.}~\bibnamefont {von~der Linde}}, \ and\ \bibinfo {author}
			{\bibfnamefont {J.~B.}\ \bibnamefont {Hastings}},\ }\href
		{https://doi.org/10.1103%2Fphysrevlett.98.125501} {\bibfield  {journal}
			{\bibinfo  {journal} {Phys. Rev. Lett.}\ }\textbf {\bibinfo {volume} {98}},\
			\bibinfo {pages} {125501} (\bibinfo {year} {2007})}\BibitemShut {NoStop}%
		\bibitem [{\citenamefont {Sciaini}\ \emph {et~al.}(2009)\citenamefont
			{Sciaini}, \citenamefont {Harb}, \citenamefont {Kruglik}, \citenamefont
			{Payer}, \citenamefont {Hebeisen}, \citenamefont {zu~Heringdorf},
			\citenamefont {Yamaguchi}, \citenamefont {von Hoegen}, \citenamefont
			{Ernstorfer},\ and\ \citenamefont {Miller}}]{Sciaini2009}%
		\BibitemOpen
		\bibfield  {author} {\bibinfo {author} {\bibfnamefont {G.}~\bibnamefont
				{Sciaini}}, \bibinfo {author} {\bibfnamefont {M.}~\bibnamefont {Harb}},
			\bibinfo {author} {\bibfnamefont {S.~G.}\ \bibnamefont {Kruglik}}, \bibinfo
			{author} {\bibfnamefont {T.}~\bibnamefont {Payer}}, \bibinfo {author}
			{\bibfnamefont {C.~T.}\ \bibnamefont {Hebeisen}}, \bibinfo {author}
			{\bibfnamefont {F.-J.~M.}\ \bibnamefont {zu~Heringdorf}}, \bibinfo {author}
			{\bibfnamefont {M.}~\bibnamefont {Yamaguchi}}, \bibinfo {author}
			{\bibfnamefont {M.~H.}\ \bibnamefont {von Hoegen}}, \bibinfo {author}
			{\bibfnamefont {R.}~\bibnamefont {Ernstorfer}}, \ and\ \bibinfo {author}
			{\bibfnamefont {R.~J.~D.}\ \bibnamefont {Miller}},\ }\href
		{https://doi.org/10.1038%2Fnature07788} {\bibfield  {journal} {\bibinfo
			{journal} {Nature}\ }\textbf {\bibinfo {volume} {458}},\ \bibinfo {pages}
		{56} (\bibinfo {year} {2009})}\BibitemShut {NoStop}%
	\bibitem [{\citenamefont {Pardini}\ \emph {et~al.}(2018)\citenamefont
		{Pardini}, \citenamefont {Alameda}, \citenamefont {Aquila}, \citenamefont
		{Boutet}, \citenamefont {Decker}, \citenamefont {Gleason}, \citenamefont
		{Guillet}, \citenamefont {Hamilton}, \citenamefont {Hayes}, \citenamefont
		{Hill}, \citenamefont {Koglin}, \citenamefont {Kozioziemski}, \citenamefont
		{Robinson}, \citenamefont {Sokolowski-Tinten}, \citenamefont {Soufli},\ and\
		\citenamefont {Hau-Riege}}]{Pardini2018}%
	\BibitemOpen
	\bibfield  {author} {\bibinfo {author} {\bibfnamefont {T.}~\bibnamefont
			{Pardini}}, \bibinfo {author} {\bibfnamefont {J.}~\bibnamefont {Alameda}},
		\bibinfo {author} {\bibfnamefont {A.}~\bibnamefont {Aquila}}, \bibinfo
		{author} {\bibfnamefont {S.}~\bibnamefont {Boutet}}, \bibinfo {author}
		{\bibfnamefont {T.}~\bibnamefont {Decker}}, \bibinfo {author} {\bibfnamefont
			{A.}~\bibnamefont {Gleason}}, \bibinfo {author} {\bibfnamefont
			{S.}~\bibnamefont {Guillet}}, \bibinfo {author} {\bibfnamefont
			{P.}~\bibnamefont {Hamilton}}, \bibinfo {author} {\bibfnamefont
			{M.}~\bibnamefont {Hayes}}, \bibinfo {author} {\bibfnamefont
			{R.}~\bibnamefont {Hill}}, \bibinfo {author} {\bibfnamefont {J.}~\bibnamefont
			{Koglin}}, \bibinfo {author} {\bibfnamefont {B.}~\bibnamefont
			{Kozioziemski}}, \bibinfo {author} {\bibfnamefont {J.}~\bibnamefont
			{Robinson}}, \bibinfo {author} {\bibfnamefont {K.}~\bibnamefont
			{Sokolowski-Tinten}}, \bibinfo {author} {\bibfnamefont {R.}~\bibnamefont
			{Soufli}}, \ and\ \bibinfo {author} {\bibfnamefont {S.}~\bibnamefont
			{Hau-Riege}},\ }\href {https://doi.org/10.1103%2Fphysrevlett.120.265701}
		{\bibfield  {journal} {\bibinfo  {journal} {Phys. Rev. Lett.}\ }\textbf
			{\bibinfo {volume} {120}},\ \bibinfo {pages} {265701} (\bibinfo {year}
			{2018})}\BibitemShut {NoStop}%
		\bibitem [{\citenamefont {Vechten}\ \emph
			{et~al.}(1979{\natexlab{a}})\citenamefont {Vechten}, \citenamefont {Tsu},
			\citenamefont {Saris},\ and\ \citenamefont {Hoonhout}}]{VanVechten1979417}%
		\BibitemOpen
		\bibfield  {author} {\bibinfo {author} {\bibfnamefont {J.~V.}\ \bibnamefont
				{Vechten}}, \bibinfo {author} {\bibfnamefont {R.}~\bibnamefont {Tsu}},
			\bibinfo {author} {\bibfnamefont {F.}~\bibnamefont {Saris}}, \ and\ \bibinfo
			{author} {\bibfnamefont {D.}~\bibnamefont {Hoonhout}},\ }\href
		{https://doi.org/10.1016%2F0375-9601%2879%2990241-x} {\bibfield  {journal}
			{\bibinfo  {journal} {Physics Letters A}\ }\textbf {\bibinfo {volume} {74}},\
			\bibinfo {pages} {417} (\bibinfo {year} {1979}{\natexlab{a}})}\BibitemShut
		{NoStop}%
		\bibitem [{\citenamefont {Vechten}\ \emph
			{et~al.}(1979{\natexlab{b}})\citenamefont {Vechten}, \citenamefont {Tsu},\
			and\ \citenamefont {Saris}}]{VANVECHTEN1979422}%
		\BibitemOpen
		\bibfield  {author} {\bibinfo {author} {\bibfnamefont {J.~V.}\ \bibnamefont
				{Vechten}}, \bibinfo {author} {\bibfnamefont {R.}~\bibnamefont {Tsu}}, \ and\
			\bibinfo {author} {\bibfnamefont {F.}~\bibnamefont {Saris}},\ }\href
		{https://doi.org/10.1016%2F0375-9601%2879%2990242-1} {\bibfield  {journal}
			{\bibinfo  {journal} {Physics Letters A}\ }\textbf {\bibinfo {volume} {74}},\
			\bibinfo {pages} {422} (\bibinfo {year} {1979}{\natexlab{b}})}\BibitemShut
		{NoStop}%
		\bibitem [{\citenamefont {Stampfli}\ and\ \citenamefont
			{Bennemann}(1994)}]{Stampfli1994}%
		\BibitemOpen
		\bibfield  {author} {\bibinfo {author} {\bibfnamefont {P.}~\bibnamefont
				{Stampfli}}\ and\ \bibinfo {author} {\bibfnamefont {K.~H.}\ \bibnamefont
				{Bennemann}},\ }\href {https://doi.org/10.1103%2Fphysrevb.49.7299} {\bibfield
			{journal} {\bibinfo  {journal} {Phys. Rev. B}\ }\textbf {\bibinfo {volume}
				{49}},\ \bibinfo {pages} {7299} (\bibinfo {year} {1994})}\BibitemShut
		{NoStop}%
		\bibitem [{\citenamefont {Zijlstra}\ \emph {et~al.}(2008)\citenamefont
			{Zijlstra}, \citenamefont {Walkenhorst}, \citenamefont {Gilfert},
			\citenamefont {Sippel}, \citenamefont {Töws},\ and\ \citenamefont
			{Garcia}}]{Zijlstra2008}%
		\BibitemOpen
		\bibfield  {author} {\bibinfo {author} {\bibfnamefont {E.~S.}\ \bibnamefont
				{Zijlstra}}, \bibinfo {author} {\bibfnamefont {J.}~\bibnamefont
				{Walkenhorst}}, \bibinfo {author} {\bibfnamefont {C.}~\bibnamefont
				{Gilfert}}, \bibinfo {author} {\bibfnamefont {C.}~\bibnamefont {Sippel}},
			\bibinfo {author} {\bibfnamefont {W.}~\bibnamefont {Töws}}, \ and\ \bibinfo
			{author} {\bibfnamefont {M.~E.}\ \bibnamefont {Garcia}},\ }\href
		{https://doi.org/10.1007%2Fs00340-008-3294-x} {\bibfield  {journal} {\bibinfo
			{journal} {Appl. Phys. B}\ }\textbf {\bibinfo {volume} {93}},\ \bibinfo
		{pages} {743} (\bibinfo {year} {2008})}\BibitemShut {NoStop}%
	\bibitem [{\citenamefont {Lian}\ \emph {et~al.}(2016)\citenamefont {Lian},
		\citenamefont {Zhang},\ and\ \citenamefont {Meng}}]{Lian2016}%
	\BibitemOpen
	\bibfield  {author} {\bibinfo {author} {\bibfnamefont {C.}~\bibnamefont
			{Lian}}, \bibinfo {author} {\bibfnamefont {S.~B.}\ \bibnamefont {Zhang}}, \
		and\ \bibinfo {author} {\bibfnamefont {S.}~\bibnamefont {Meng}},\ }\href
	{https://doi.org/10.1103%2Fphysrevb.94.184310} {\bibfield  {journal}
		{\bibinfo  {journal} {Phys. Rev. B}\ }\textbf {\bibinfo {volume} {94}},\
		\bibinfo {pages} {184310} (\bibinfo {year} {2016})}\BibitemShut {NoStop}%
	\bibitem [{\citenamefont {Stampfli}\ and\ \citenamefont
		{Bennemann}(1992)}]{Stampfli1992}%
	\BibitemOpen
	\bibfield  {author} {\bibinfo {author} {\bibfnamefont {P.}~\bibnamefont
			{Stampfli}}\ and\ \bibinfo {author} {\bibfnamefont {K.~H.}\ \bibnamefont
			{Bennemann}},\ }\href {https://doi.org/10.1103%2Fphysrevb.46.10686}
		{\bibfield  {journal} {\bibinfo  {journal} {Phys. Rev. B}\ }\textbf {\bibinfo
				{volume} {46}},\ \bibinfo {pages} {10686} (\bibinfo {year}
			{1992})}\BibitemShut {NoStop}%
		\bibitem [{\citenamefont {Yabana}\ and\ \citenamefont
			{Bertsch}(1996)}]{Yabana1996}%
		\BibitemOpen
		\bibfield  {author} {\bibinfo {author} {\bibfnamefont {K.}~\bibnamefont
				{Yabana}}\ and\ \bibinfo {author} {\bibfnamefont {G.~F.}\ \bibnamefont
				{Bertsch}},\ }\href {https://doi.org/10.1103%2Fphysrevb.54.4484} {\bibfield
			{journal} {\bibinfo  {journal} {Phys. Rev. B}\ }\textbf {\bibinfo {volume}
				{54}},\ \bibinfo {pages} {4484} (\bibinfo {year} {1996})}\BibitemShut
		{NoStop}%
		\bibitem [{\citenamefont {Marques}(2003)}]{Marques2003}%
		\BibitemOpen
		\bibfield  {author} {\bibinfo {author} {\bibfnamefont {M.}~\bibnamefont
				{Marques}},\ }\href {https://doi.org/10.1016%2Fs0010-4655%2802%2900686-0}
			{\bibfield  {journal} {\bibinfo  {journal} {Computer Physics Communications}\
				}\textbf {\bibinfo {volume} {151}},\ \bibinfo {pages} {60} (\bibinfo {year}
				{2003})}\BibitemShut {NoStop}%
			\bibitem [{\citenamefont {Castro}\ \emph {et~al.}(2006)\citenamefont {Castro},
				\citenamefont {Appel}, \citenamefont {Oliveira}, \citenamefont {Rozzi},
				\citenamefont {Andrade}, \citenamefont {Lorenzen}, \citenamefont {Marques},
				\citenamefont {Gross},\ and\ \citenamefont {Rubio}}]{Castro2006}%
			\BibitemOpen
			\bibfield  {author} {\bibinfo {author} {\bibfnamefont {A.}~\bibnamefont
					{Castro}}, \bibinfo {author} {\bibfnamefont {H.}~\bibnamefont {Appel}},
				\bibinfo {author} {\bibfnamefont {M.}~\bibnamefont {Oliveira}}, \bibinfo
				{author} {\bibfnamefont {C.~A.}\ \bibnamefont {Rozzi}}, \bibinfo {author}
				{\bibfnamefont {X.}~\bibnamefont {Andrade}}, \bibinfo {author} {\bibfnamefont
					{F.}~\bibnamefont {Lorenzen}}, \bibinfo {author} {\bibfnamefont {M.~A.~L.}\
					\bibnamefont {Marques}}, \bibinfo {author} {\bibfnamefont {E.~K.~U.}\
					\bibnamefont {Gross}}, \ and\ \bibinfo {author} {\bibfnamefont
					{A.}~\bibnamefont {Rubio}},\ }\href
			{https://doi.org/10.1002%2Fpssb.200642067} {\bibfield  {journal} {\bibinfo
				{journal} {phys. stat. sol. (b)}\ }\textbf {\bibinfo {volume} {243}},\
			\bibinfo {pages} {2465} (\bibinfo {year} {2006})}\BibitemShut {NoStop}%
		\bibitem [{\citenamefont {Krieger}\ \emph {et~al.}(2015)\citenamefont
			{Krieger}, \citenamefont {Dewhurst}, \citenamefont {Elliott}, \citenamefont
			{Sharma},\ and\ \citenamefont {Gross}}]{Krieger2015}%
		\BibitemOpen
		\bibfield  {author} {\bibinfo {author} {\bibfnamefont {K.}~\bibnamefont
				{Krieger}}, \bibinfo {author} {\bibfnamefont {J.~K.}\ \bibnamefont
				{Dewhurst}}, \bibinfo {author} {\bibfnamefont {P.}~\bibnamefont {Elliott}},
			\bibinfo {author} {\bibfnamefont {S.}~\bibnamefont {Sharma}}, \ and\ \bibinfo
			{author} {\bibfnamefont {E.~K.~U.}\ \bibnamefont {Gross}},\ }\href
		{https://doi.org/10.1021%2Facs.jctc.5b00621} {\bibfield  {journal} {\bibinfo
			{journal} {J. Chem. Theory Comput.}\ }\textbf {\bibinfo {volume} {11}},\
		\bibinfo {pages} {4870} (\bibinfo {year} {2015})}\BibitemShut {NoStop}%
	\bibitem [{\citenamefont {Elliott}\ \emph {et~al.}(2016)\citenamefont
		{Elliott}, \citenamefont {Krieger}, \citenamefont {Dewhurst}, \citenamefont
		{Sharma},\ and\ \citenamefont {Gross}}]{Elliott2016}%
	\BibitemOpen
	\bibfield  {author} {\bibinfo {author} {\bibfnamefont {P.}~\bibnamefont
			{Elliott}}, \bibinfo {author} {\bibfnamefont {K.}~\bibnamefont {Krieger}},
		\bibinfo {author} {\bibfnamefont {J.~K.}\ \bibnamefont {Dewhurst}}, \bibinfo
		{author} {\bibfnamefont {S.}~\bibnamefont {Sharma}}, \ and\ \bibinfo {author}
		{\bibfnamefont {E.~K.~U.}\ \bibnamefont {Gross}},\ }\href
	{https://doi.org/10.1088%2F1367-2630%2F18%2F1%2F013014} {\bibfield  {journal}
		{\bibinfo  {journal} {New J. Phys.}\ }\textbf {\bibinfo {volume} {18}},\
		\bibinfo {pages} {013014} (\bibinfo {year} {2016})}\BibitemShut {NoStop}%
	\bibitem [{\citenamefont {Andrade}\ \emph {et~al.}(2015)\citenamefont
		{Andrade}, \citenamefont {Strubbe}, \citenamefont {Giovannini}, \citenamefont
		{Larsen}, \citenamefont {Oliveira}, \citenamefont {Alberdi-Rodriguez},
		\citenamefont {Varas}, \citenamefont {Theophilou}, \citenamefont {Helbig},
		\citenamefont {Verstraete}, \citenamefont {Stella}, \citenamefont {Nogueira},
		\citenamefont {Aspuru-Guzik}, \citenamefont {Castro}, \citenamefont
		{Marques},\ and\ \citenamefont {Rubio}}]{Andrade2015}%
	\BibitemOpen
	\bibfield  {author} {\bibinfo {author} {\bibfnamefont {X.}~\bibnamefont
			{Andrade}}, \bibinfo {author} {\bibfnamefont {D.}~\bibnamefont {Strubbe}},
		\bibinfo {author} {\bibfnamefont {U.~D.}\ \bibnamefont {Giovannini}},
		\bibinfo {author} {\bibfnamefont {A.~H.}\ \bibnamefont {Larsen}}, \bibinfo
		{author} {\bibfnamefont {M.~J.~T.}\ \bibnamefont {Oliveira}}, \bibinfo
		{author} {\bibfnamefont {J.}~\bibnamefont {Alberdi-Rodriguez}}, \bibinfo
		{author} {\bibfnamefont {A.}~\bibnamefont {Varas}}, \bibinfo {author}
		{\bibfnamefont {I.}~\bibnamefont {Theophilou}}, \bibinfo {author}
		{\bibfnamefont {N.}~\bibnamefont {Helbig}}, \bibinfo {author} {\bibfnamefont
			{M.~J.}\ \bibnamefont {Verstraete}}, \bibinfo {author} {\bibfnamefont
			{L.}~\bibnamefont {Stella}}, \bibinfo {author} {\bibfnamefont
			{F.}~\bibnamefont {Nogueira}}, \bibinfo {author} {\bibfnamefont
			{A.}~\bibnamefont {Aspuru-Guzik}}, \bibinfo {author} {\bibfnamefont
			{A.}~\bibnamefont {Castro}}, \bibinfo {author} {\bibfnamefont {M.~A.~L.}\
			\bibnamefont {Marques}}, \ and\ \bibinfo {author} {\bibfnamefont
			{A.}~\bibnamefont {Rubio}},\ }\href {https://doi.org/10.1039%2Fc5cp00351b}
		{\bibfield  {journal} {\bibinfo  {journal} {Phys. Chem. Chem. Phys.}\
			}\textbf {\bibinfo {volume} {17}},\ \bibinfo {pages} {31371} (\bibinfo {year}
			{2015})}\BibitemShut {NoStop}%
		\bibitem [{\citenamefont {Meng}\ \emph {et~al.}(2008)\citenamefont {Meng},
			\citenamefont {Ren},\ and\ \citenamefont {Kaxiras}}]{Meng2008}%
		\BibitemOpen
		\bibfield  {author} {\bibinfo {author} {\bibfnamefont {S.}~\bibnamefont
				{Meng}}, \bibinfo {author} {\bibfnamefont {J.}~\bibnamefont {Ren}}, \ and\
			\bibinfo {author} {\bibfnamefont {E.}~\bibnamefont {Kaxiras}},\ }\href
		{https://doi.org/10.1021%2Fnl801644d} {\bibfield  {journal} {\bibinfo
			{journal} {Nano Lett.}\ }\textbf {\bibinfo {volume} {8}},\ \bibinfo {pages}
		{3266} (\bibinfo {year} {2008})}\BibitemShut {NoStop}%
	\bibitem [{\citenamefont {Lian}\ \emph
		{et~al.}(2018{\natexlab{a}})\citenamefont {Lian}, \citenamefont {Guan},
		\citenamefont {Hu}, \citenamefont {Zhang},\ and\ \citenamefont
		{Meng}}]{Lian2018}%
	\BibitemOpen
	\bibfield  {author} {\bibinfo {author} {\bibfnamefont {C.}~\bibnamefont
			{Lian}}, \bibinfo {author} {\bibfnamefont {M.}~\bibnamefont {Guan}}, \bibinfo
		{author} {\bibfnamefont {S.}~\bibnamefont {Hu}}, \bibinfo {author}
		{\bibfnamefont {J.}~\bibnamefont {Zhang}}, \ and\ \bibinfo {author}
		{\bibfnamefont {S.}~\bibnamefont {Meng}},\ }\href
	{https://doi.org/10.1002%2Fadts.201800055} {\bibfield  {journal} {\bibinfo
		{journal} {Adv. Theory Simul.}\ }\textbf {\bibinfo {volume} {1}},\ \bibinfo
	{pages} {1800055} (\bibinfo {year} {2018}{\natexlab{a}})}\BibitemShut
{NoStop}%
\bibitem [{\citenamefont {Lian}\ \emph
	{et~al.}(2018{\natexlab{b}})\citenamefont {Lian}, \citenamefont {Hu},
	\citenamefont {Guan},\ and\ \citenamefont {Meng}}]{Lian2018b}%
\BibitemOpen
\bibfield  {author} {\bibinfo {author} {\bibfnamefont {C.}~\bibnamefont
		{Lian}}, \bibinfo {author} {\bibfnamefont {S.-Q.}\ \bibnamefont {Hu}},
	\bibinfo {author} {\bibfnamefont {M.-X.}\ \bibnamefont {Guan}}, \ and\
	\bibinfo {author} {\bibfnamefont {S.}~\bibnamefont {Meng}},\ }\href
{https://doi.org/10.1063%2F1.5036543} {\bibfield  {journal} {\bibinfo
	{journal} {The Journal of Chemical Physics}\ }\textbf {\bibinfo {volume}
	{149}},\ \bibinfo {pages} {154104} (\bibinfo {year}
{2018}{\natexlab{b}})}\BibitemShut {NoStop}%
\bibitem [{\citenamefont {Ordej{\'{o}}n}\ \emph {et~al.}(1996)\citenamefont
	{Ordej{\'{o}}n}, \citenamefont {Artacho},\ and\ \citenamefont
	{Soler}}]{Ordejon1996}%
\BibitemOpen
\bibfield  {author} {\bibinfo {author} {\bibfnamefont {P.}~\bibnamefont
		{Ordej{\'{o}}n}}, \bibinfo {author} {\bibfnamefont {E.}~\bibnamefont
		{Artacho}}, \ and\ \bibinfo {author} {\bibfnamefont {J.~M.}\ \bibnamefont
		{Soler}},\ }\href {https://doi.org/10.1103%2Fphysrevb.53.r10441} {\bibfield
	{journal} {\bibinfo  {journal} {Phys. Rev. B}\ }\textbf {\bibinfo {volume}
		{53}},\ \bibinfo {pages} {R10441} (\bibinfo {year} {1996})}\BibitemShut
{NoStop}%
\bibitem [{\citenamefont {Soler}\ \emph {et~al.}(2002)\citenamefont {Soler},
	\citenamefont {Artacho}, \citenamefont {Gale}, \citenamefont {Garc{\'{\i}}a},
	\citenamefont {Junquera}, \citenamefont {Ordej{\'{o}}n},\ and\ \citenamefont
	{S{\'{a}}nchez-Portal}}]{Soler2002}%
\BibitemOpen
\bibfield  {author} {\bibinfo {author} {\bibfnamefont {J.~M.}\ \bibnamefont
		{Soler}}, \bibinfo {author} {\bibfnamefont {E.}~\bibnamefont {Artacho}},
	\bibinfo {author} {\bibfnamefont {J.~D.}\ \bibnamefont {Gale}}, \bibinfo
	{author} {\bibfnamefont {A.}~\bibnamefont {Garc{\'{\i}}a}}, \bibinfo {author}
	{\bibfnamefont {J.}~\bibnamefont {Junquera}}, \bibinfo {author}
	{\bibfnamefont {P.}~\bibnamefont {Ordej{\'{o}}n}}, \ and\ \bibinfo {author}
	{\bibfnamefont {D.}~\bibnamefont {S{\'{a}}nchez-Portal}},\ }\href
{https://doi.org/10.1088%2F0953-8984%2F14%2F11%2F302} {\bibfield  {journal}
	{\bibinfo  {journal} {J. Phys.: Condens. Matter}\ }\textbf {\bibinfo {volume}
		{14}},\ \bibinfo {pages} {2745} (\bibinfo {year} {2002})}\BibitemShut
{NoStop}%
\bibitem [{\citenamefont {{S{\'{a}}nchez-Portal, Daniel and Ordej{\'{o}}n,
			Pablo and Artacho, Emilio and Soler}}\ \emph {et~al.}(1997)\citenamefont
	{{S{\'{a}}nchez-Portal, Daniel and Ordej{\'{o}}n, Pablo and Artacho, Emilio
			and Soler}}, \citenamefont {S{\'{a}}nchez-Portal}, \citenamefont
	{Ordej{\'{o}}n}, \citenamefont {Artacho},\ and\ \citenamefont
	{Soler}}]{Sanchez-Portal1997}%
\BibitemOpen
\bibfield  {author} {\bibinfo {author} {\bibfnamefont {J.~M.}\ \bibnamefont
		{{S{\'{a}}nchez-Portal, Daniel and Ordej{\'{o}}n, Pablo and Artacho, Emilio
				and Soler}}}, \bibinfo {author} {\bibfnamefont {D.}~\bibnamefont
		{S{\'{a}}nchez-Portal}}, \bibinfo {author} {\bibfnamefont {P.}~\bibnamefont
		{Ordej{\'{o}}n}}, \bibinfo {author} {\bibfnamefont {E.}~\bibnamefont
		{Artacho}}, \ and\ \bibinfo {author} {\bibfnamefont {J.~M.}\ \bibnamefont
		{Soler}},\ }\href {\doibase
	10.1002/(SICI)1097-461X(1997)65:5<453::AID-QUA9>3.0.CO;2-V} {\bibfield
	{journal} {\bibinfo  {journal} {International Journal of Quantum Chemistry}\
	}\textbf {\bibinfo {volume} {65}},\ \bibinfo {pages} {453} (\bibinfo {year}
	{1997})}\BibitemShut {NoStop}%
\bibitem [{\citenamefont {Troullier}\ and\ \citenamefont
	{Martins}(1991)}]{Troullier1991}%
\BibitemOpen
\bibfield  {author} {\bibinfo {author} {\bibfnamefont {N.}~\bibnamefont
		{Troullier}}\ and\ \bibinfo {author} {\bibfnamefont {J.~L.}\ \bibnamefont
		{Martins}},\ }\href {https://doi.org/10.1103%2Fphysrevb.43.8861} {\bibfield
	{journal} {\bibinfo  {journal} {Phys. Rev. B}\ }\textbf {\bibinfo {volume}
		{43}},\ \bibinfo {pages} {8861} (\bibinfo {year} {1991})}\BibitemShut
{NoStop}%
\bibitem [{\citenamefont {Perdew}\ and\ \citenamefont
	{Zunger}(1981)}]{Perdew1981}%
\BibitemOpen
\bibfield  {author} {\bibinfo {author} {\bibfnamefont {J.~P.}\ \bibnamefont
		{Perdew}}\ and\ \bibinfo {author} {\bibfnamefont {A.}~\bibnamefont
		{Zunger}},\ }\href {https://doi.org/10.1103%2Fphysrevb.23.5048} {\bibfield
	{journal} {\bibinfo  {journal} {Phys. Rev. B}\ }\textbf {\bibinfo {volume}
		{23}},\ \bibinfo {pages} {5048} (\bibinfo {year} {1981})}\BibitemShut
{NoStop}%
\bibitem [{\citenamefont {Harb}\ \emph {et~al.}(2008)\citenamefont {Harb},
	\citenamefont {Ernstorfer}, \citenamefont {Hebeisen}, \citenamefont
	{Sciaini}, \citenamefont {Peng}, \citenamefont {Dartigalongue}, \citenamefont
	{Eriksson}, \citenamefont {Lagally}, \citenamefont {Kruglik},\ and\
	\citenamefont {Miller}}]{Harb2008}%
\BibitemOpen
\bibfield  {author} {\bibinfo {author} {\bibfnamefont {M.}~\bibnamefont
		{Harb}}, \bibinfo {author} {\bibfnamefont {R.}~\bibnamefont {Ernstorfer}},
	\bibinfo {author} {\bibfnamefont {C.~T.}\ \bibnamefont {Hebeisen}}, \bibinfo
	{author} {\bibfnamefont {G.}~\bibnamefont {Sciaini}}, \bibinfo {author}
	{\bibfnamefont {W.}~\bibnamefont {Peng}}, \bibinfo {author} {\bibfnamefont
		{T.}~\bibnamefont {Dartigalongue}}, \bibinfo {author} {\bibfnamefont {M.~A.}\
		\bibnamefont {Eriksson}}, \bibinfo {author} {\bibfnamefont {M.~G.}\
		\bibnamefont {Lagally}}, \bibinfo {author} {\bibfnamefont {S.~G.}\
		\bibnamefont {Kruglik}}, \ and\ \bibinfo {author} {\bibfnamefont {R.~J.~D.}\
		\bibnamefont {Miller}},\ }\href
{https://doi.org/10.1103%2Fphysrevlett.100.155504} {\bibfield  {journal}
	{\bibinfo  {journal} {Phys. Rev. Lett.}\ }\textbf {\bibinfo {volume} {100}},\
	\bibinfo {pages} {155504} (\bibinfo {year} {2008})}\BibitemShut {NoStop}%
\bibitem [{\citenamefont {Dronskowski}\ and\ \citenamefont
	{Bloechl}(1993)}]{Dronskowski1993}%
\BibitemOpen
\bibfield  {author} {\bibinfo {author} {\bibfnamefont {R.}~\bibnamefont
		{Dronskowski}}\ and\ \bibinfo {author} {\bibfnamefont {P.~E.}\ \bibnamefont
		{Bloechl}},\ }\href {https://doi.org/10.1021%2Fj100135a014} {\bibfield
	{journal} {\bibinfo  {journal} {J. Phys. Chem.}\ }\textbf {\bibinfo {volume}
		{97}},\ \bibinfo {pages} {8617} (\bibinfo {year} {1993})}\BibitemShut
{NoStop}%
\bibitem [{\citenamefont {Rohringer}\ \emph {et~al.}(2006)\citenamefont
	{Rohringer}, \citenamefont {Peter},\ and\ \citenamefont
	{Burgdörfer}}]{Rohringer2006a}%
\BibitemOpen
\bibfield  {author} {\bibinfo {author} {\bibfnamefont {N.}~\bibnamefont
		{Rohringer}}, \bibinfo {author} {\bibfnamefont {S.}~\bibnamefont {Peter}}, \
	and\ \bibinfo {author} {\bibfnamefont {J.}~\bibnamefont {Burgdörfer}},\
}\href {https://doi.org/10.1103%2Fphysreva.74.042512} {\bibfield  {journal}
{\bibinfo  {journal} {Phys. Rev. A}\ }\textbf {\bibinfo {volume} {74}},\
\bibinfo {pages} {042512} (\bibinfo {year} {2006})}\BibitemShut {NoStop}%
\end{thebibliography}%
\end{document}